\documentclass[referee]{aa}
\pdfoutput=1
\usepackage{graphicx}
\usepackage{epstopdf}
\usepackage{txfonts}
\usepackage{natbib}
\bibpunct{(}{)}{;}{a}{}{,}
\usepackage{longtable}
\usepackage{url}
\usepackage{lscape}
\begin{document}
   \title{VOSA: Virtual Observatory SED Analyzer.}

   \subtitle{An application to the Collinder 69 open cluster}

     \author{A. Bayo
          \inst{1,2}
          \and
	  C. Rodrigo
	  \inst{1,2}
	  \and
          D. Barrado y Navascu\'es
	  \inst{1,2}
	  \and
	  E. Solano
	  \inst{1,2}
	  \and
	  R. Guti\'errez
	  \inst{1,2}
	  \and
	  M. Morales--Calder\'on
	  \inst{1}
	  \and
	  F. Allard
	  \inst{3}
          }

   \offprints{A. Bayo\\ 
   \email{abayo@laeff.inta.es}}

   \institute{Laboratorio de Astrof\'{\i}sica Espacial y F\'{\i}sica Fundamental (LAEFF-CAB, INTA-CSIC), P.O. Box 78, 28691 Villanueva de la Ca\~{n}ada, Madrid, Spain
     \and
     SVO Thematic Network, Spain
     \and
     Centre de Recherche Astronomique de Lyon (CRAL), \'Ecole Normale Sup\'erieure de Lyon, 69364, Lyon, France
   }

   \date{Received ...; accepted ...}

 
  \abstract
  {The physical properties of almost any kind of astronomical object can be derived by fitting synthetic spectra or photometry extracted from 
theoretical models to observational data.}
  {We want to develop an automatic procedure to perform this kind of fittings to a relatively large sample of members of a stellar association and apply this methodology to the case of Collinder 69.}
{We combine the multiwavelength data of our sources and follow a work-flow to derive the physical parameters of the sources. The key step of the work-flow is performed by a new VO-tool, VOSA. All the steps in this process are done in a VO environment.}
  {We present this new tool, and provide physical parameters such as T$_{\rm eff}$, gravity, luminosity, etc. for $\sim$170 candidate members to Collinder 69, and an upper-limit for the age of this stellar association.} 
{This kind of studies of star forming regions, clusters, etc. produces a huge amount of data, 
  very tedious to analyse using the traditional methodology. Thus, they are excellent examples where to apply the VO capabilities.}
   \titlerunning{Virtual Observatory SED Analizer. The case of Collinder 69}
   \maketitle
%

\section{Introduction}
\label{intro}

The determination of physical parameters of astronomical objects from observational data
is frequently linked with the use of theoretical models as templates. In the case of stars,
it is very common to estimate gravities by fitting different model/templates to gravity--sensitive lines 
(and other features, \citealt{Decin04}) within the wavelength coverage of the observed spectra.
Another method in the same framework is the comparison of the shape of the Spectral 
Energy Distributions (SEDs hereafter) constructed from observational photometric data with the corresponding synthetic ones derived using large grids of theoretical spectra (see for example \citealt{Robitaille07}). 

The use, -in the traditional way-, of these methodologies can easily become tedious and even unfeasible when applied to large amount 
of data. Nowadays, astronomers deal with very large databases, including not only empirical/observational entries, but also theoretical ones (for instance, models from di\-ffe\-rent groups that need to be combined in order to reproduce the observations).

The Virtual Observatory (VO) philosophy (see Section~\ref{TVOT}) seems to be a reasonable approach to solve some intrinsic problems generated by this new way of working in astronomy. 
VO-tools\footnote{ http://ivoa.net/twiki/bin/IVOA/IvoaApplications} 
now allow astronomers to deal with their own observational or theoretical databases and compare them (in an efficient way) with others already published in the ``on-line literature''.

Here we present a work-flow to estimate the effective temperature (T$_{\rm eff}$), gravity and luminosity (and, when comparing with theoretical 
isochrones and evolutionary tracks, age and mass) of members of an stellar association (in this particular case, Collinder 69, \citealt{Barrado04}) based on their
photometry. This work-flow has been built mainly with steps that are ``VO-compliant''. 

In Section~\ref{SC} we 
explain the scientific case that caused the development of VOSA\footnote{http://www.laeff.inta.es/svo/theory/vosa/}, the key VO-tool used in this work-flow. In Section~\ref{TVOT} we describe the philosophy
of the Virtual Observatory and some of its most known applications (tools that we have used in several steps of the 
process). In that section we also explain all the capabilities of the previously mentioned VO-tool VOSA (developed by the Spanish Virtual Observatory\footnote{ https://laeff.inta.es/svo/} specially for this work).
In Section~\ref{TM} we explain step by step the work-flow followed to determine the physical parameters and properties of our sources.
In Sections ~\ref{Results} and ~\ref{C}, we present the results and conclusions of this work, respectively.
Finally, in Appendix A, we focus on technical issues concerning the computation of synthetic photometry  and in Appendix B we provide a graphical scheme of the work-flow.


\section{The scientific case}
\label{SC}

One of the most interesting Star Forming Regions (SFR) is associated to the O8III star $\lambda$~Orionis (located at about 400 pc from the Sun, and presenting very low extinction in its inner area, a central ring where A$_V \sim 0.36$ magnitudes). This star dominates the eponymous cluster (also designated as Collinder 69), with an age of about 5 Myr \citep{DM99,DM01,Barrado04}. 

The $\lambda$~Orionis SFR includes several distinct associations apart from Collinder 69 (as Barnard 30, Barnard 35, and dark clouds like LDN 1588 and 1603). But, for this work, we will focus on the central cluster. Our goal is to determine physical parameters and properties (such as T$_{\rm eff}$, bolometric luminosity, mass and age) of $\sim$ 170 candidate members of this cluster by comparing their SEDs with theoretical models. 

Studying the physical parameters of a large population of sources belonging to the same cluster is advantageous, since we can infer properties not only of the individual sources but also of the association as a whole, for example its age, since we can assume that all objects are coeval.
In this work we want to derive these physical parameters and properties by comparing observed SEDs with theoretical data; therefore, as a first step, we need to compile all the photometric/spectral information available for each of our sources in order to be able to decide which model reproduces best the observed data. 

On the other hand, there are different collections of models (translated in thousands of individual models) that we might want to compare to the observations. 

Trying to compile all the information available for each object and then compare the built SED with all the available models one by one with classical methodologies can be very tedious and time consuming, therefore we need to automatise the process as much as possible.


\section{The VO--tools}
\label{TVOT}

The Virtual Observatory (IVOA\footnote{http://www.ivoa.net}) is an international, community-based initiative to 
provide seamless access to the data available from astronomical archives and services as well as state-of-the-art 
tools for the efficient analysis of this huge amount of information. The works by \citet{Padovani04} and 
\citet{Tsalmantza06} are good examples of the efficiency of such tools in helping astronomers 
to produce scientific results.

\subsection{General-purpose tools}
In this work we have taken advantage of the capabilities of Aladin\footnote{http://aladin.u-strasbg.fr/aladin.gml}, an 
interactive sky atlas developed by CDS that allows the user to visualise and analyse astronomical images, spectra and 
catalogues available from the VO services. We have also used TOPCAT\footnote{http://www.star.bris.ac.uk/~mbt/topcat/}, an interactive 
graphical viewer and editor for tabular data as well as different tools for analysing astronomical spectra 
(VOSED\footnote{ http://sdc.laeff.inta.es/vosed/}, VOSpec\footnote{ http://esavo.esa.int/vospec}, 
SPECVIEW\footnote{ http://www.stsci.edu/resources/software\_hardware/ specview/users} and 
SPLAT\footnote{ http://www.starlink.ac.uk/splat}).

Regarding the data format, we have used the standard tabular format of the VO environment, the VOTable\footnote{ http://www.ivoa.net/Documents/latest/VOT.html}, whenever possible.

\subsection{VOSA}
\label{VOSA}

As mentioned in the Introduction, a new VO-tool has been developed by the Spanish Virtual Observatory for the scientific case discussed in Section~\ref{SC}. This tool has been named VOSA (Virtual Observatory SED Analyzer), and can be accessed through its web-base interface at:\\
\url{http://www.laeff.inta.es/svo/theory/vosa/}

In order to use VOSA, the user only has to register in the service and start working. The system is based on sessions that can be saved, so that the user can recover the files and results obtained in a previous session (without redoing the fit process) and compare those results with new approaches (models, parameters, etc.). These sessions will remain in the system unless there is no activity in seven days (in that case the session will be erased).

VOSA performs the following tasks:
\begin{enumerate}
\item Read user photometry-tables. 

The input file can be either an ASCII table or a VOTable but with a specific format consisting of eight columns/fields as follows: 

{\bf Source Identifier}: this will be used, for instance, as the title of the plots of the fittings. 

{\bf Coordinates of the source}: Right ascension and Declination (in J2000) will be used to look for counterparts in various catalogs.

{\bf Distance to the source in parsecs}: this datum is mandatory in order to estimate the bolometric luminosity of the source. Should this parameter be unknown then type in a value of ``$---$" and  a generic distance of 10 parsecs will be assumed.

{\bf Visual extincion} (A$_{\rm V}$ in magnitudes) affecting the source: this value is used to deredden each SED (the \citealt{Indebetouw05} extinction law is applied).

{\bf Filter name}: a list of the already available filters can be found in the Help menu, new filters can be included upon user request. 

{\bf Observed flux/magnitude} (flux in units of erg/cm$^2$/s/\AA). 

{\bf Flux/magnitude error} (same units as in the previous field). 

For an example of the data ingestion interface see Fig.~\ref{fig:input1} in Appendix B. A more detailed description of the file format, available filters, fluxes for zero magnitude, etc. can be accessed from the VOSA Help menu.

\item Query several photometric catalogs accessible through VO services in order to increase the wavelength coverage of the data to be analised. Nowadays the user can query the 2MASS All-Sky Point Source Catalog \citep{2MASS}, the Tycho-2 Catalog \citep{Tycho},  and the Str\"{o}mgren uvby-beta Catalog  \citep{Stromgren} using different radii according to the astrometic precision of each catalog.

\item Query VO-compliant theoretical models (spectra) for a given range of physical parameters.

The user can choose among the different available collections of models (in a future implementation of the tool, it will offer the possibility to access any VO-compliant theoretical model collection). 
These models are accessible in a VO-environment from the SVO theoretical data server at \url{http://laeff.inta.es/svo/theory/db2vo/}

\item Calculate the synthetic photometry of the theoretical spectra (within the range of physical parameters required by the user) for the set of filters used to obtain the observed data (including the dataset coming from the VO photometrical services). 

The synthetic photometry is calculated by convolving the filter response curve with the synthetic spectra. Prior to this, both, the response curve and the synthetic spectra, are interpolated to match the spectral resolution; besides, the area of the response curve is normalised to unity to provide the user with a flux density that can be directly compared with the observed one (see Appendix A for the formal description of the process).

Since some of the response curves for the set of filters that we use in our scientific case (for example the four channels of Spitzer/IRAC, \citealt{Fazio04}, and the 24 micron channel of Spitzer/MIPS, \citealt{Rieke04}) were not available in the VO Filter Profile 
Service\footnote{ http://voservices.net/filter/filterlist.aspx ?mode=keyword\&keyword=}, we consulted directly the web pages of the respective consortia and included them in VOSA (the transmission curves can be visualised and downloaded from the Help menu).  

\item Perform a statistical test to decide which set of synthetic photometry reproduces best the observed data. 

The provided ``best" fitting model is the one that minimises the value of the reduced $\chi^{2}$ defined as: 
\begin{equation}
\chi^{2} = \frac {1}{N-P} \sum \left\{ \frac{1}{\sigma_0^2}(Y_0-M_{\rm d} \times Y_{\rm m})^2\right\}
\end{equation}
 where $Y_0$ is the observed flux, $\sigma _0$ the observational error in the flux, $N$ the number of photometric points, $P$ the number of parameters being fitted, $Y_{\rm m}$ the theoretical flux predicted by the model and $M_d$ the multiplicative dilution factor, defined as $(R/D)^2$ (for models from \citealt{Hauschildt99, Allard01} and \citealt{Chabrier00}), $R$ being the radius of the source and $D$ the distance to the object.

During the fitting process, the tool detects possible infrared excesses (in our scientific case some of the sources might be surrounded by disks). Thus, since the theoretical spectra correspond to stellar atmospheres, for the calculation of the $\chi^2$ the tool only considers those data points of the SED corresponding to bluer wavelengths than the one where the excess has been flagged. 

The excesses are detected by calculating iteratively (adding a new data point from the SED at a time), in the mid-infrared (wavelengths redder than 2.5$\mu$m), the $\alpha$ parameter as defined in \citet{Lada06} (which becomes larger than -2.56 when the source present an infrared excess). The last wavelength considered in the fitting process together with the ratio between the total number of points belonging to the SED to those really used are displayed in the result tables.

Once the fitting process is completed, the tool allows the user to check whether any other of the five ``best" fits, per collection of models (meaning 20 best fits when using the four collections of models available in the server nowadays), turns out to be more convincing for him/her than the one that minimises the $\chi^2$ and choose that one as the best model to include in the final results table. For instance, the user might have a more accurate estimation of the gravity, which is important since changes in the surface gravity of stellar and sub-stellar sources do not produce dramatic differences in the shape of their SEDs.

\item Use the best-fit model as the source of a bolometric correction. 

The best fitting model is used to infer the total observed flux for each source of the sample {\bf ; this process is performed by integrating the total observed flux (where no excess has been detected), and using the best fitting model to infer that flux in the wavelength range where no observational data is available or some infrared excess has been detected} (see Appendix A for a formal description of the procedure). We note that if the model reproduces the data correctly, this estimation is much more accurate than the one obtained using a bolometric correction derived only from a single colour.

{\bf Regarding the estimation of the error in the calculated ``total observed flux"; VOSA uses the model to infer the total flux emitted by the source in the wavelength range where no observational data is provided  or some infrared excess has been detected; since we cannot provide any uncertainties for the model itself we have extrapolated the errors of the observed photometry to the whole wavelength range, and therefore we provide an error for the total flux parameter dependent on the accuracy of the provided measurements.}

\item Provide the estimated bolometric luminosity for each object. 

The tool scales the total observed flux to the distance given by the user and therefore estimates the bolometric luminosity of each source in the sample. In those cases where the user has not provided a value of the distance, a generic value of 10 parsecs is assumed:
\begin{equation}
\label{eq:lum}
L = 4 \pi d^{2} F_{\rm Obs}
\end{equation}

\item Generate a Hertzsprung-Russell diagram with the estimated parameters. 

The tool provides the user with the possibility to select and overplot in this diagram different sets of isochrones and evolutionary tracks with masses and ages within the range selected by him/her.

\item Provide an estimation of the mass and age of the individual sources. 

VOSA interpolates among the previously mentioned collections of isochrones and evolutionary tracks to estimate the mass and age of each source from the input sample (see the Help menu from VOSA for details on the interpolation processes and possible warning flags).

\end{enumerate}

As can be easily derived from the previous description, even though VOSA has been designed to fulfill the requirements of our scientific case in Collinder 69, this tool performs various tasks that are commonly used in different fields of astronomy in a completely automatic manner. Therefore, we expect that a significant part of the astronomical community will find VOSA a very useful tool. 

\section{The case of Collinder 69: Goals and methodology}
\label{TM}

Our goals are the following:
\begin{itemize}

\item Build the Spectral Energy Distribution for every object in our sample using the photometric (and even spectroscopical) data 
available in astronomical archives and services (a typical VO task).

\item Determine physical properties (such as effective temperature, surface gravity and luminosity) of each object of interest by comparing its SED with those derived from theoretical spectra.

Since our sources are members of a young cluster and we are fitting models of photospheres, we will have to deal with the infrared excesses that might be present in some of the sources.

\item Compare the physical parameters obtained with theoretical isochrones and evolutionary tracks in order
to estimate the age and mass of each individual target of the sample, and the age of the association 
as a whole.
\end{itemize}

We have followed the steps described previously using only VO-compliant tools. 
Some of the tools were already available, others (as VOSA, and the plotting capabilities of the 
theoretical isochrones and evolutionary tracks server) have been specifically developed for this particular case but can be applied in a much more general frame. 

\subsection{The Data}

For the determination of the physical parameters of the members of Collinder 69 we have made use of optical to mid-infrared photometry. The data have been taken from \citet{Barrado07} and complemented with VO photometric services (see Section~\ref{VOSA} for details). None of our sources had Tycho or Hauck/Str\"{o}mgren counterparts, but some of them did have 2MASS counterparts (we have used 2MASS photometry for those sources for which we did not have near IR photometry either from CAHA/Omega2000 or WHT/INGRID, see \citealt{Barrado07} for details). Ten photometric points per object have been typically used in the analysis (see an example in Table~\ref{table1}). 

\begin{table*}
\caption{Example of the photometric data for our sources.} 
\label{table1}
\begin{scriptsize}
\begin{tabular}{|l|c|c|c|c|c|c|c|c|c|c|}
\hline
Name&R$^{\mathrm{a}}$&I$^{\mathrm{a}}$&J$^{\mathrm{b}}$&H$^{\mathrm{b}}$&Ks$^{\mathrm{b}}$&[3.6]$^{\mathrm{c}}$&[4.5]$^{\mathrm{c}}$&[5.8]$^{\mathrm{c}}$&[8.0]$^{\mathrm{c}}$&[24]$^{\mathrm{d}}$\\
\hline
\hline
\noalign{\smallskip}
...     & ...            & ...            & ...              & ...              & ...              & ...              & ...               & ...              & ...              & ...  \\
LOri029 & 14.89$\pm$0.00 & 13.69$\pm$0.00 & 12.210$\pm$0.026 & 11.460$\pm$0.027 & 11.071$\pm$0.019 & 10.259$\pm$0.003 & 9.830$\pm$0.003   & 9.321$\pm$0.006  & 8.416$\pm$0.003  & 5.684$\pm$0.007 \\ 
LOri034 & 15.10$\pm$0.00 & 13.97$\pm$0.00 & 12.442$\pm$0.026 & 11.639$\pm$0.026 & 11.184$\pm$0.023 & 10.068$\pm$0.003 & 9.734$\pm$0.003   & 9.314$\pm$0.007  & 8.325$\pm$0.003  & 5.738$\pm$0.007 \\ 
...     & ...            & ...            & ...              & ...              & ...              & ...              & ...               & ...              & ...              & ...  \\
\hline
\end{tabular}
\end{scriptsize}
\vspace*{0.2cm}

\begin{footnotesize}
Filters corresponding to the telescope/instrument configuration described below:\\
$^{\mathrm{a}}$CFHT\\
$^{\mathrm{b}}$2MASS, \citet{2MASS}\\
$^{\mathrm{c}}$Spitzer/IRAC, \citet{Fazio04}\\
$^{\mathrm{d}}$Spitzer/MIPS, \citet{Rieke04}\\
\end{footnotesize}
\end{table*}

\subsection{The Work-flow}

\begin{figure}
\resizebox{\hsize}{!}{\includegraphics{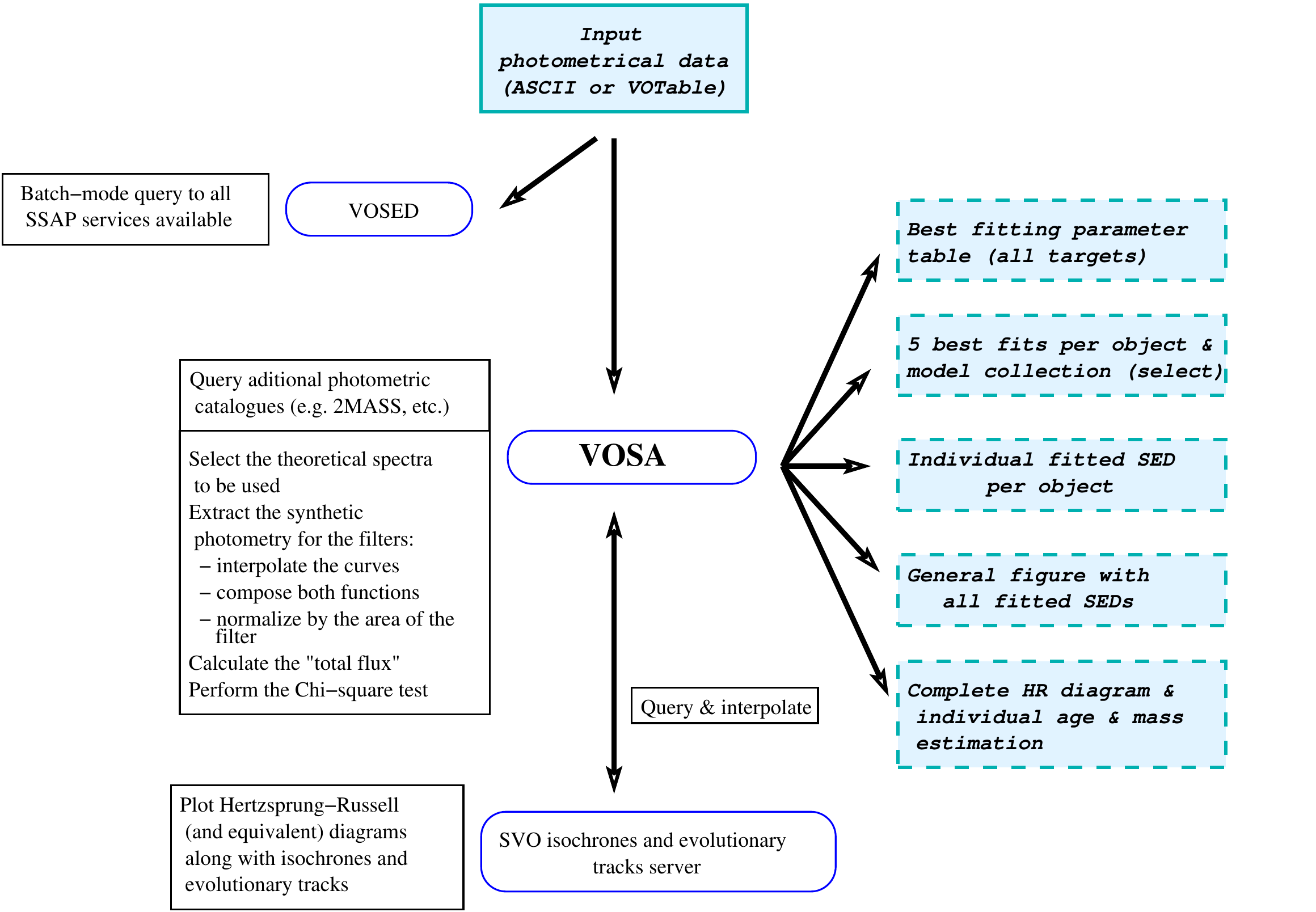}}
\caption{Work-flow scheme. The shaded squared represent the input data (top) and the products obtained when the workflow 
has been completed (bottom-right corner, dashed box). The names of the applications used are written in rounded boxes, and the tasks performed by
these applications are listed in rectangular boxes on the left hand side.}
\label{fig:WF}
\end{figure}

\begin{figure}
\resizebox{\hsize}{!}{\includegraphics{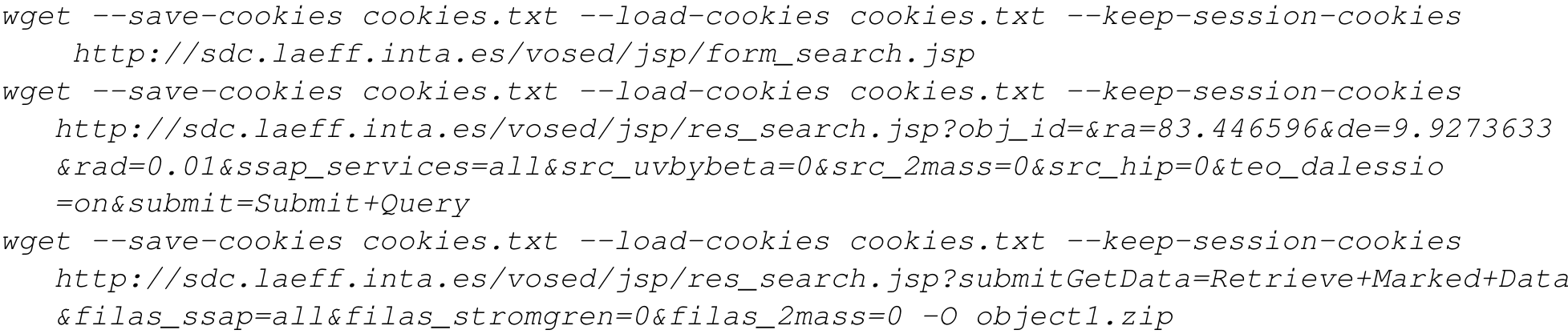}}
\caption{Example of the three lines per object to include in the script. 
With the first line, the user access the VOSED form; with the second one, a query is performed (to the services available 
in the VOSED form) around the position written in the fields {\it ra} and {\it de} within the radius {\it rad} (in degrees). 
Finally with the third line the available data for those coordinates is retrieved.
All the information in saved in one ``zip-file''.}
\label{fig:script}
\end{figure}

\begin{figure}
\resizebox{\hsize}{!}{\includegraphics{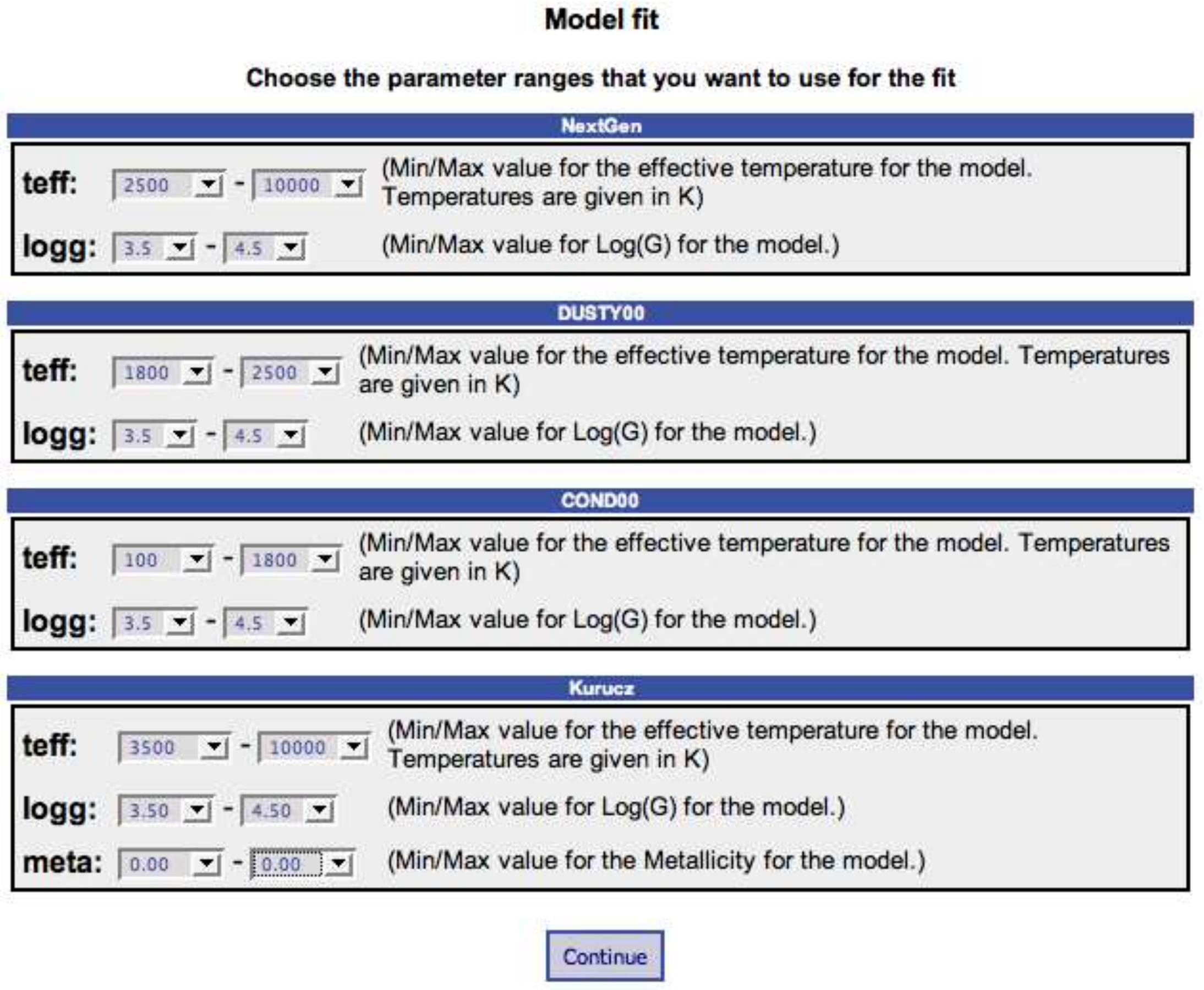}}
\caption{Range of parameters queried for each collection of synthetic spectra.}
\label{ParamRange}
\end{figure}

\begin{figure}
\resizebox{\hsize}{!}{\includegraphics{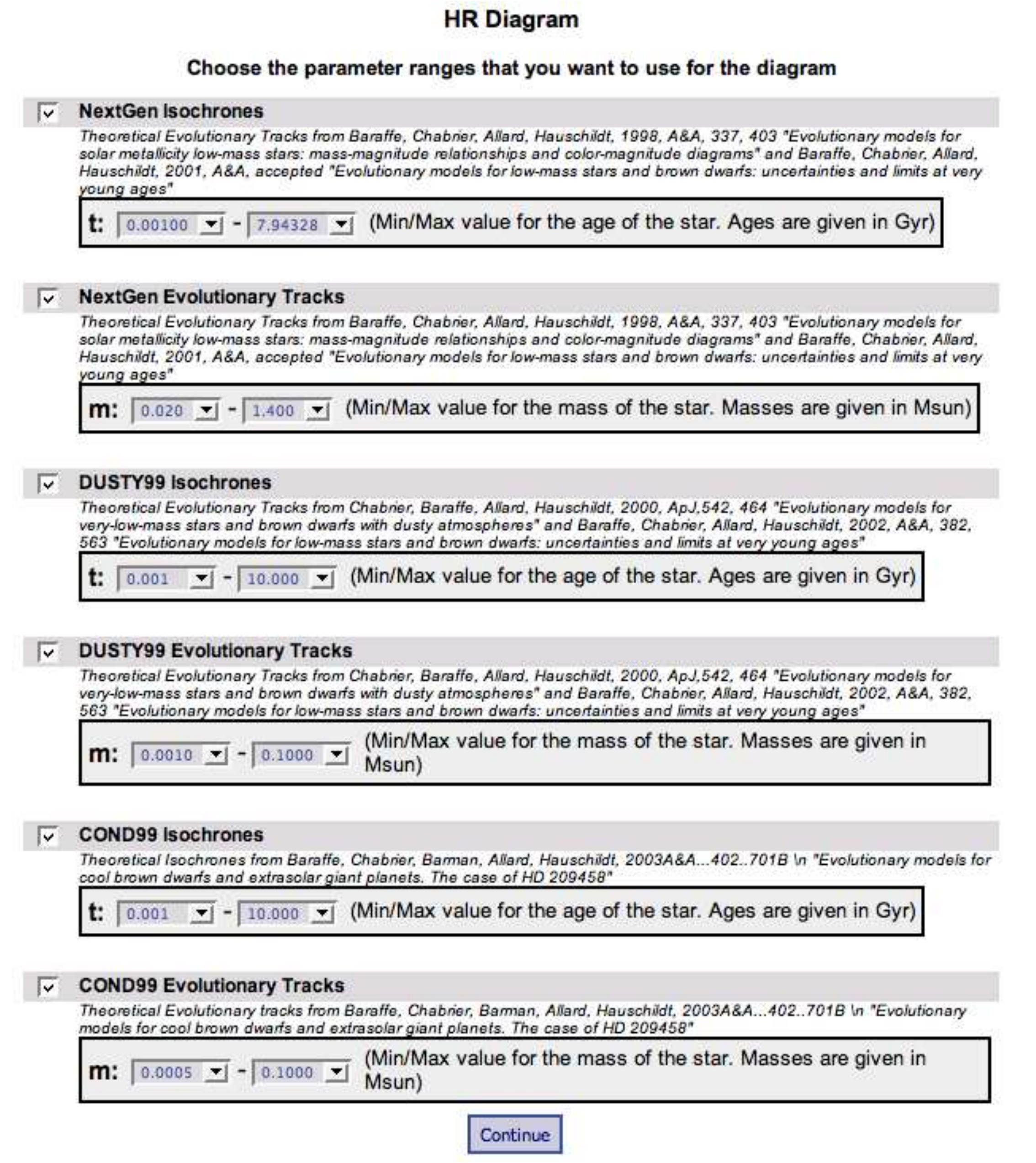}}
\caption{Range of parameters queried for each collection of isochrones and evolutionary tracks.}
\label{fig:isochr_form}
\end{figure}

\begin{figure}
\resizebox{\hsize}{!}{\includegraphics{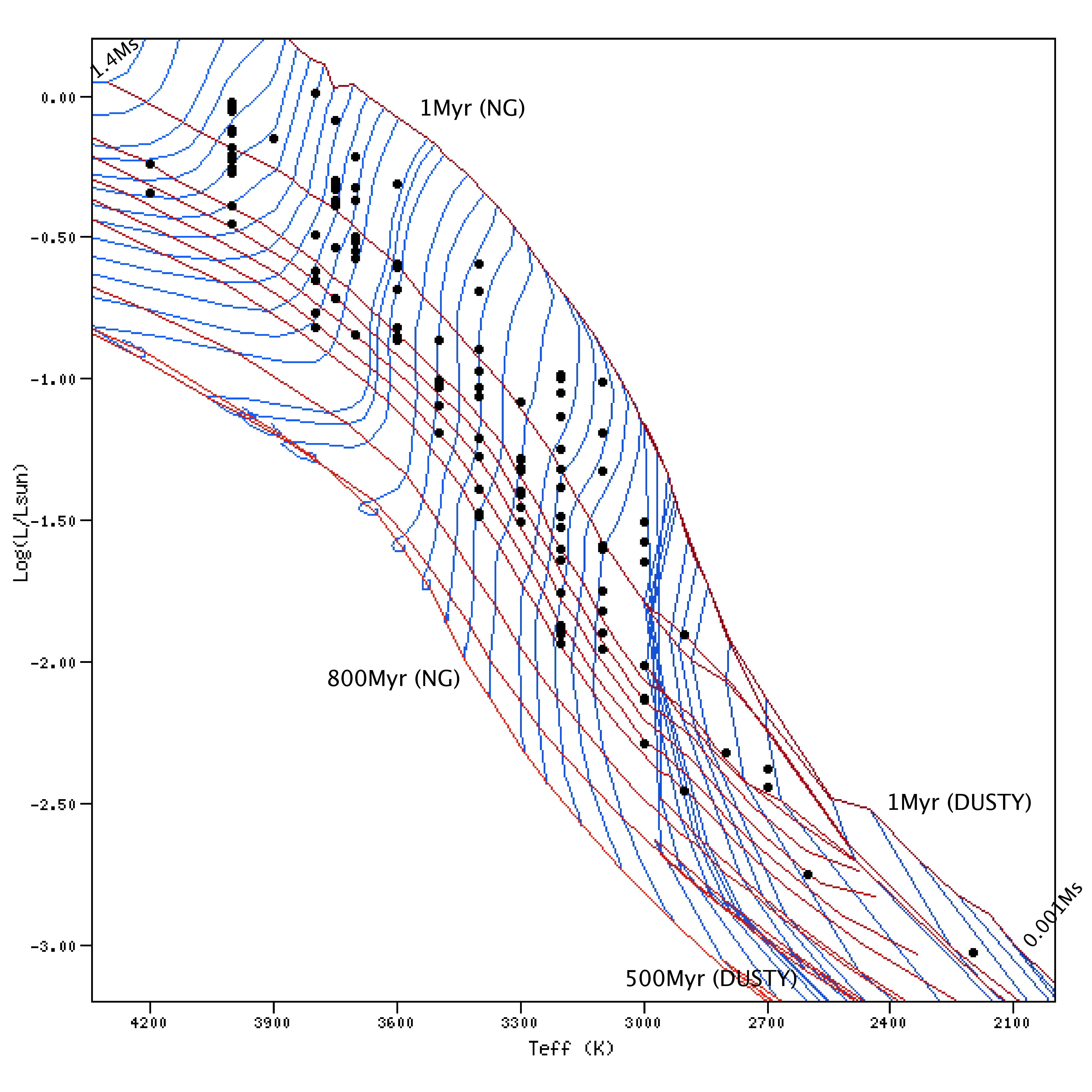}}
\caption{HR diagram of the members of Collinder 69 for which no infrared excess was detected. Isochrones corresponding to ages of 1, 5, 10, 12.5, 16, 20, 25, 50, 100 and 800 Myrs are displayed for the NextGen collection, and those corresponding to ages of 1, 5, 10, 50, 100,120 and 500 Myr for the DUSTY collection. Evolutionary tracks are also displayed for masses between 0.001M\sun~and 1.4M\sun~(including both, NextGen and DUSTY collections).}
\label{fig:HR}
\end{figure}

We will now describe the workflow fo\-llo\-wed to achieve our goals (see the scheme in Fig.~\ref{fig:WF} and Appendix B for a more detailed visual workflow):

\begin{itemize}

\item Build the observed SEDs. This was accomplished in two steps:

\begin{enumerate}

\item VOSA\footnote{http://www.laeff.inta.es/svo/theory/vosa/} was used to get Tycho, Str\"{o}mgren and 2MASS counterparts within given search radii (see Fig.~\ref{VOphot2} in Appendix B)

\item VOSED 
was queried to look for spectroscopic observations from VO services. Given the number of objects to query around (167), we prepared a script to make this query in batch-mode (see Fig.~\ref{fig:script}, Appendix B). No spectroscopic information via VO-servers was available for any of the sources.

\end{enumerate}

The so-constructed observational SED can be visualized with different VO-tools, namely, VOSpec
, Specview 
and SPLAT.


\item Determine effective temperatures, surface gravities and estimate the total observed flux.

This task was performed completely using the new VO-tool VOSA (described in detail in Section~\ref{VOSA}). The following steps were taken:

\begin{enumerate}

\item Select the theoretical models (see Fig~\ref{ParamRange} for details on the range of parameters queried). At present, only  Kurucz (ODFNEW/NOVER models, \citealt{Castelli97}), NEXTGEN \citep{Hauschildt99}, DUSTY \citep{Allard01} and COND \citep{Chabrier00} models are available. This is enough for our scientific case (due to the expected properties of our sources), but as mentioned before, it is our aim to expand the number of models by giving access to all the collections available at the SVO theoretical data sever and, ultimately, to the rest of VO services. 

\item Absolute flux calibration of the observational data. As mentioned before, all the information regarding the available filters (zero points, transmission curves, etc.) can be consulted through the help menu of VOSA.

\item Determine the synthetic photometry for the models with physical parameters in the range selected by the user (in this case, different ranges of T$_{\rm eff}$ depending on the characteristics of the models, and $\log g$ between 3.5 and 4.5 dex, due to the expected youth of Collinder 69).

\item Model fitting: the best fitting model (as explained in Section~\ref{VOSA}) is provided by VOSA.

\end{enumerate}

VOSA provides the user with different outputs (see Fig.~\ref{fig:WF}, for a schematic view). For each object, the five models (per collection of models) that best fit the observations are displayed. Tabular (i.e. physical parameters) and graphical (i.e. plots) results are provided for each fit. The system also allows the user to take advantage of his/her own knowledge of the objects and select the actual best among any of the displayed 20 fittings.

Once a definite set of best fits its achieved, the tool offers different types of output files containing information about the fitting: 

\begin{enumerate}

\item A summary table (all objects in ASCII or VOTable format) for each entry in the input catalogue with the following fields: identifier, theoretical model collection, effective temperature, surface gravity, reduced-$\chi^2$ and multiplicative dilution factor (factor to scale the theoretical data to the observed one; in the case of the models form the Lyon group is related to the ratio between the radius and the distance to the source) of the best fitting, number of data points considered for the calculation of the $\chi^2$, the effective wavelength corresponding to the filter of the reddest data-point considered in the fitting process (a flag indicating whether the source has an infrared excess or not), estimated total observed flux, the bolometric luminosity 
 and the ratio of total estimated flux by the total observed flux (a measurement of the bolometric correction applied). Links to the observational photometry as well as to the synthetic photometry and spectra are also provided in this table (see Fig.~\ref{bigtable} from Appendix B).


\item An overall figure with the best fits for all the entries in the input catalog (see Fig.~\ref{big_plot} from Appendix B)    

\end{enumerate}


\item Determination of ages and masses.

This task was performed by combining the capabilities of the the\-o\-re\-ti\-cal isochrones and evolutionary tracks service created by the SVO (available at: \url{http://laeff.inta.es/svo/theory/draw/getiso.php?inises=guest}) and VOSA. The process is the following:
\begin{enumerate}
\item Access the HR diagram menu, and select the option to create a new diagram.
\item Decide what collection/s of isochrones and evolutionary tracks (and what ranges of ages and masses) should be used in the interpolation process (in our case we queried the whole range of possibilities, see Fig.~\ref{fig:isochr_form} from Appendix B). Nowadays the available collections are the following: NextGen isochrones and evolutionary tracks from \citet{Baraffe98}, DUSTY isochrones and evolutionary tracks from \citet{Chabrier00} and COND isochrones and evolutionary tracks from \citet{Baraffe03}

We must note that to perform the interpolation for each particular source, VOSA will use the curves (both isochrones and evolutionary tracks) corresponding to the collection of models for which the best fit, -or the one flagged by the user as best-, was achieved. Otherwise the consistency of the method cannot be guaranteed.

\item Check the estimations given by VOSA (different flags explain possible sources of problems in the estimation process; see the Help menu for further details), and ``play" with the sets of isochrones and evolutionary tracks, that should be displayed in order to clarify the HR diagram produced by the application (see as an example Fig~\ref{fig:HR} from Appendix B, where most of the sources are displayed in an HR diagram, and different sets of isochrones and evolutionary tracks are overplotted). 
\end{enumerate}

\end{itemize}

Once all these estimations are available, we are able to study the cluster as a whole by the statistics derived for the sample of sources, and also study each object individually. The results derived from this study are presented in the next section.


\section{Collinder 69 as an example: results}
\label{Results}
\subsection{General outputs}
\label{GResults}

Following the previously described method we have been able to build Spectral Energy Distributions and estimate effective temperatures, gravities and bolometric luminosities for 167 candidate members of Collinder 69, and even infer ages and masses for 145 of these sources in an {\bf automatic manner}.

\begin{figure}
\resizebox{\hsize}{!}{\includegraphics{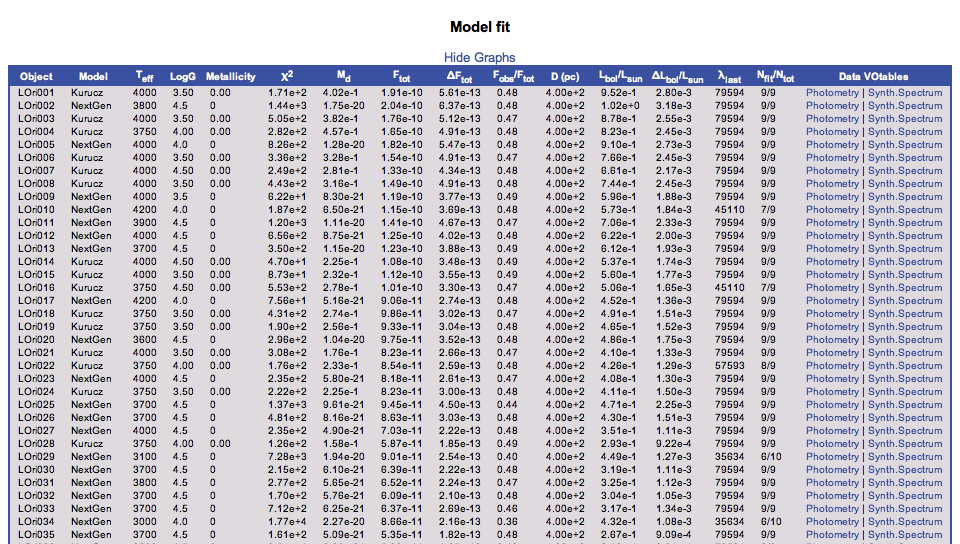}}
\caption{Several rows of the ``master-table'' with all the fittings.}
\label{bigtable}
\end{figure}

\begin{figure}
\resizebox{\hsize}{!}{\includegraphics{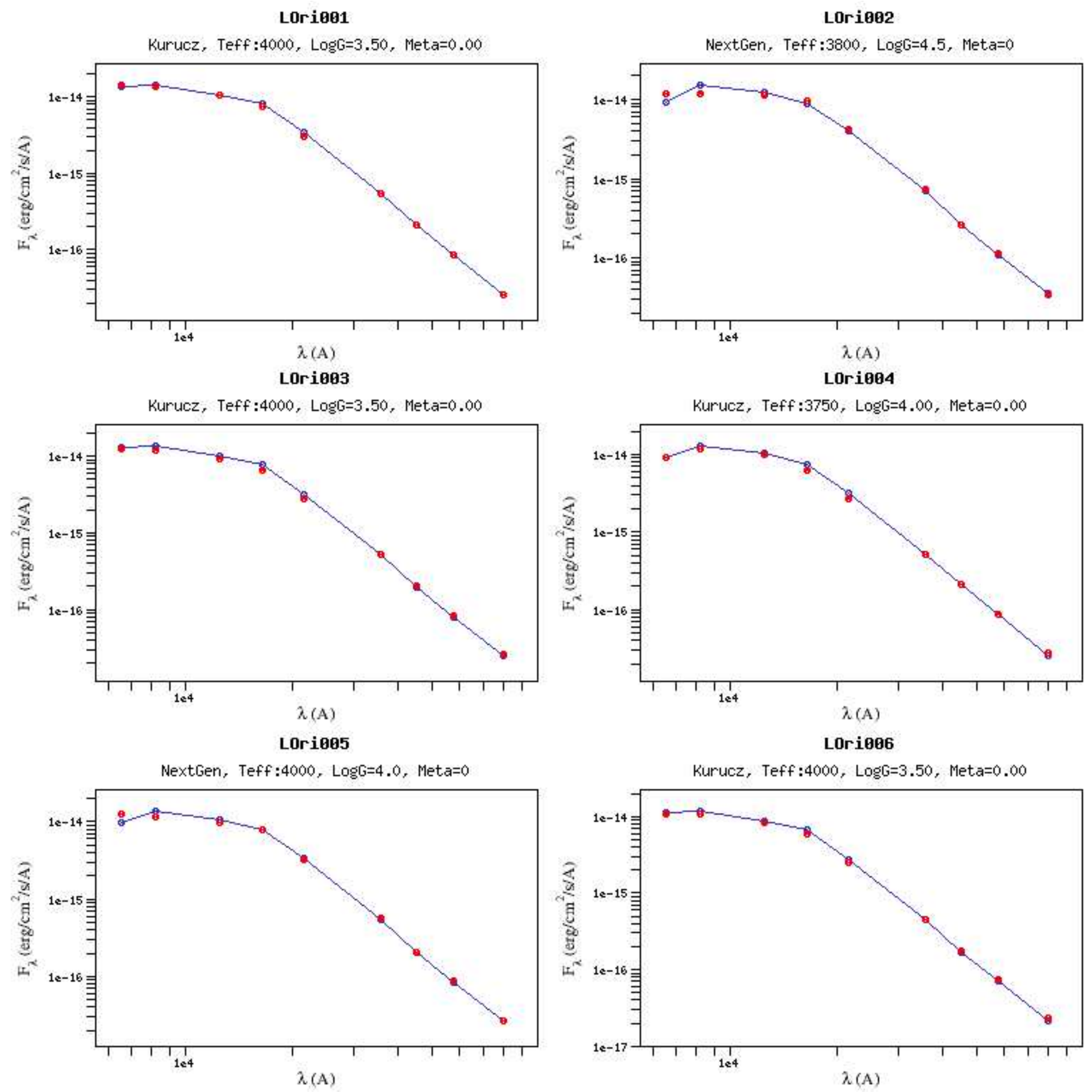}}
\caption{First page from the graphical results output.}
\label{big_plot}
\end{figure}

\begin{figure}
\resizebox{\hsize}{!}{\includegraphics{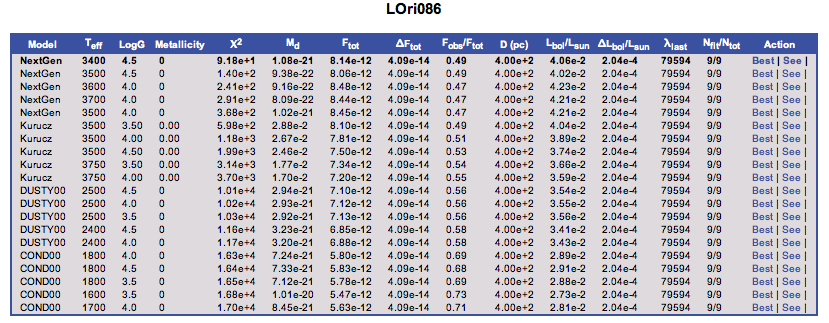}}
\caption{Example of the ``small table'' showing the 15 best fittings for an individual object (LOri086).}
\label{fig:output2}
\end{figure}

\begin{figure}
\resizebox{\hsize}{!}{\includegraphics{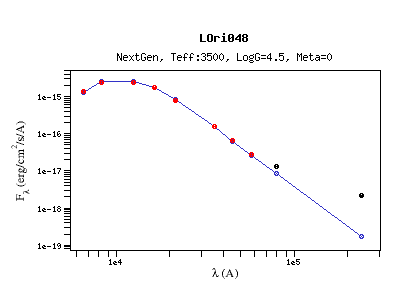}}
\caption{Graphical output from VOSA for a source with infrared excess (LOri048). The black dots highlight those data points not considered when performing the fitting process.}
\label{fig:output3}
\end{figure}

Examples of the estimations made through synthetic photometry fittings are shown in Figs.~\ref{bigtable},~\ref{big_plot},~\ref{fig:output2} and~\ref{fig:output3} and the results achieved for each source 
are shown in Table~\ref{all_param}.

Regarding the comparison with theoretical isochrones and evolutionary tracks in the Hertzsprung-Russell diagram, 
VOSA does not provide estimations of either age or mass for 20 sources ($\sim$12\% from the total) due to the peculiar location of these sources in that diagram (see VOSA Help menu for further details on the cases where the interpolation process cannot be completed and no estimation of age and/or mass can be provided). 

\begin{figure}
\resizebox{\hsize}{!}{\includegraphics{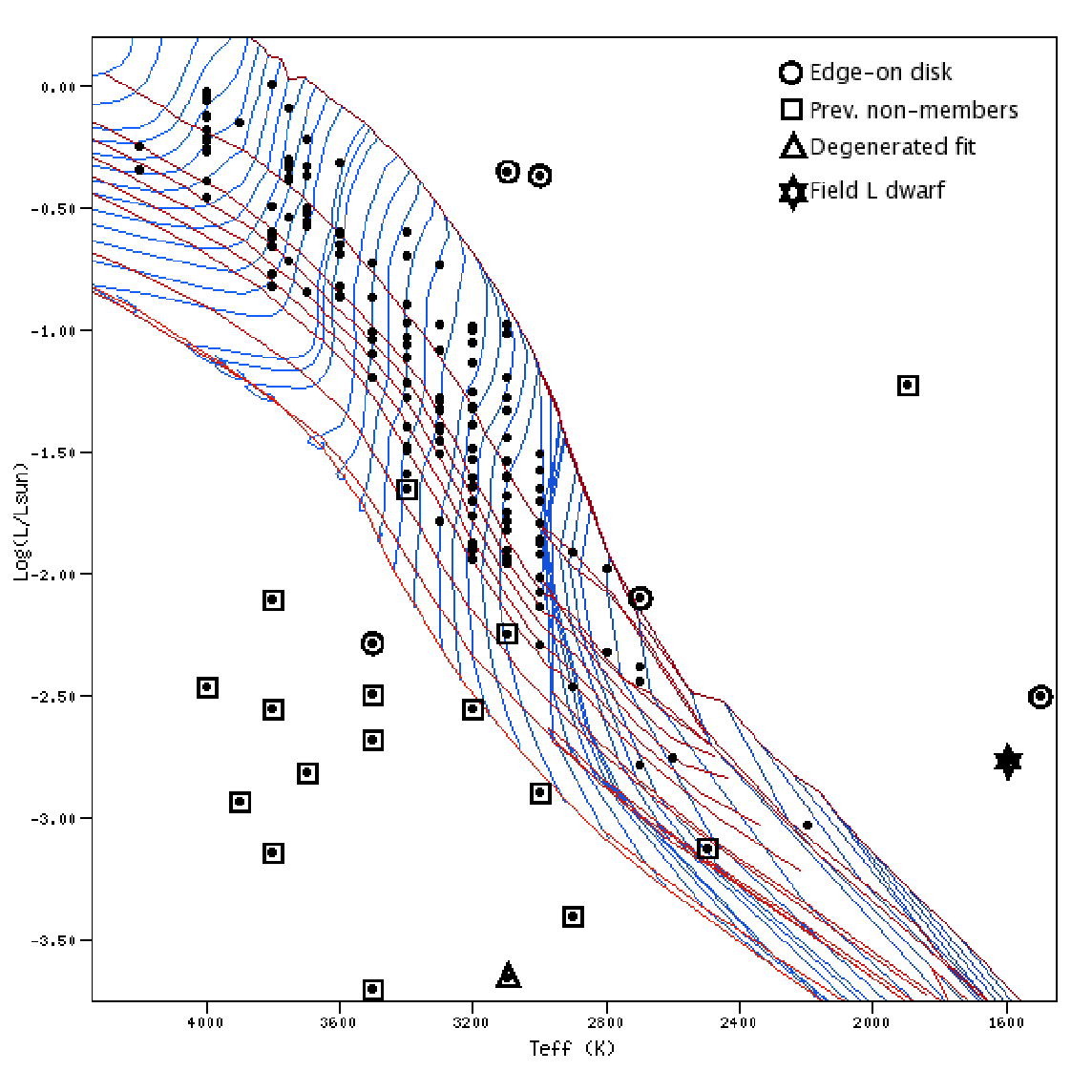}}
\caption{Location in a HR diagram of the ``peculiar sources" analyzed in Section~\ref{IResults}. Squares surround non-members candidates from \citet{Barrado07}; circles are displayed around sources showing infrared excess, and a triangle and a star highlight LOri169 and LOri162, respectively. See text for details.}
\label{fig:HRall}
\end{figure}

Besides the case of two special sources, there are two main causes for this peculiar location in the HR diagram: either the source might harbour a close to edge-on disk that reprocess the light emitted from the photosphere of the source (and therefore both, the effective temperature and the bolometric luminosity are underestimated) or the source does not belong to cluster. See Fig.~\ref{fig:HRall} for details on the specific location of the peculiar sources and Subsection~\ref{IResults} for the discussion on each particular case.

{\bf 
\subsection{Comparison with other methodologies.}
In order to test the results provided by VOSA, we have estimated some of the parameters and properties discussed in this work with those derived using other methodologies (mainly based on colours and classical bolometric corrections).

\subsubsection{Infrared excesses.}
\citet{Barrado07} studied in detail the Spitzer/IRAC data of the candidates, and based on the criteria developed by \citet{Allen04} and \citet{Hartmann05} derived a fraction of members with disks of 25\% and 40\% for the stellar and substellar population of C69, respectively. 

VOSA detects infrared excesses by calculating iteratively the $\alpha$ parameter from \citet{Lada06}. This method provides us with excess fractions of:  32\% for candidates with stellar masses (from 0.72~M$\odot$ to 1.15~M$\odot$ ), and 44\% for sources with estimated masses $\le 0.72$~M$\odot$. The differences can be explained by the presence of optically thin disks that mimic the location of single photospheres in the Spitzer/IRAC colour-colour diagrams and are therefore classified as Class III sources.

\subsubsection{Effective temperature estimation.}

We have derived the effective temperature of our candidates using two different colours ([R-I] and [I-Ks]) and temperature-colour relationships (\citealt{Bessell91} and \citealt{Leggett92}). In Fig.~\ref{teff_comp} we have plotted the relationships mentioned before and overplotted as solid and open circles (diskless and disk harbouring sources) the results obtained with VOSA. \\
There is a good agreement among the different relationships (specially with the one proposed by \citealt{Bessell91}) with offsets of -150K and +100K for the two colours, [R-I] and [I-Ks], respectively. These offsets have already been reported in the literature when comparing temperatures obtained with models and those obtained with colours and temperature scales (see for example \citealt{Barrado04a}, \citealt{Barrado06} and \citealt{Mohanty07}).\\
We have selected one specific case, namely LOri055, where the differences between the temperatures derived from the colors and the one estimated by VOSA are much higher than the mentioned offset. 
For this case we have forced VOSA to find the best multiplicative factor to fit the SED of this object with NextGen models at temperatures in the range between the lowest (3200K from the [I-Ks] color and the \citealt{Leggett92} temperature scale) and the highest (3600K from VOSA) derived for the source.\\
In Fig~\ref{LOri055} we show the fits obtained when forcing the different temperatures. An inspection of the behaviour of the $\chi^2$ of the fits concludes that the temperature proposed by VOSA improves one and a half orders of magnitude the goodness of the fit obtained with the lowest temperature (the one derived with the [I-Ks] colour).

\begin{figure}
\resizebox{\hsize}{!}{\includegraphics{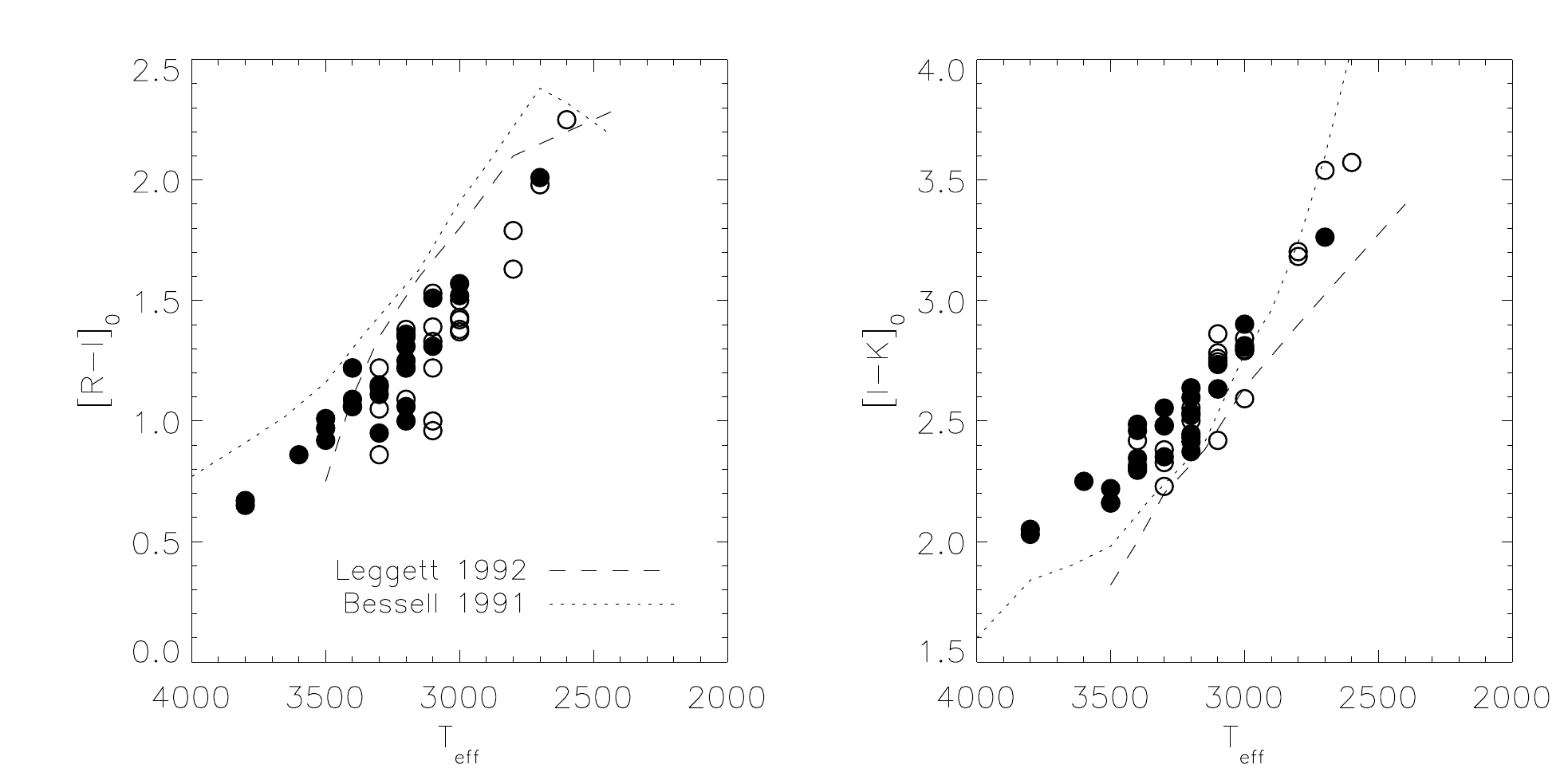}}
\caption{Deredened colors CFHT [R-I] and CFHT--2MASS [I-K] vs. T$_{\rm eff}$ (obtained with VOSA) for the candidate members of C69. The dotted line represents the relationship derived by \citet{Bessell91} and the dashed line the one derived by \citet{Leggett92}. Open circles highlight sources harbouring disks.}
\label{teff_comp}
\end{figure}

\begin{figure}
\resizebox{\hsize}{!}{\includegraphics{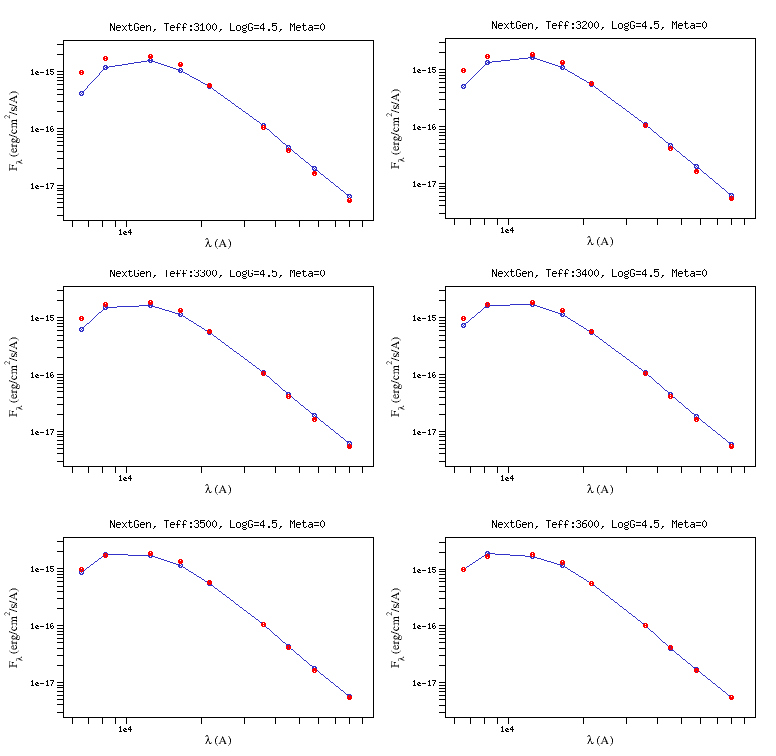}}
\caption{Forced fits to the SED of LOri055 for the range of effective temperatures estimated according to the different methods. According to our calculations, the $\chi^2$ improves one and a half orders of magnitude when comparing the 3100K and the 3600K fits.}
\label{LOri055}
\end{figure}

\subsubsection{Bolometric luminosity estimation.}

In the case of the computation of the bolometric luminosity, we have applied different bolometric corrections to several photometric bands (I band by \citealt{Kenyon95}, J band by \citealt{Lawson96} and Ks band by \citealt{Tinney93}) and as can be seen in Fig.~\ref{llum_comp}, there are no systematic differences among the various methods. We must note however, than in our case we use the whole observed SED (but for the wavelength range where some excess is detected) to infer the total emitted flux, and therefore, the better the accuracy of the observations, the better the estimation.

\begin{figure}
\resizebox{\hsize}{!}{\includegraphics{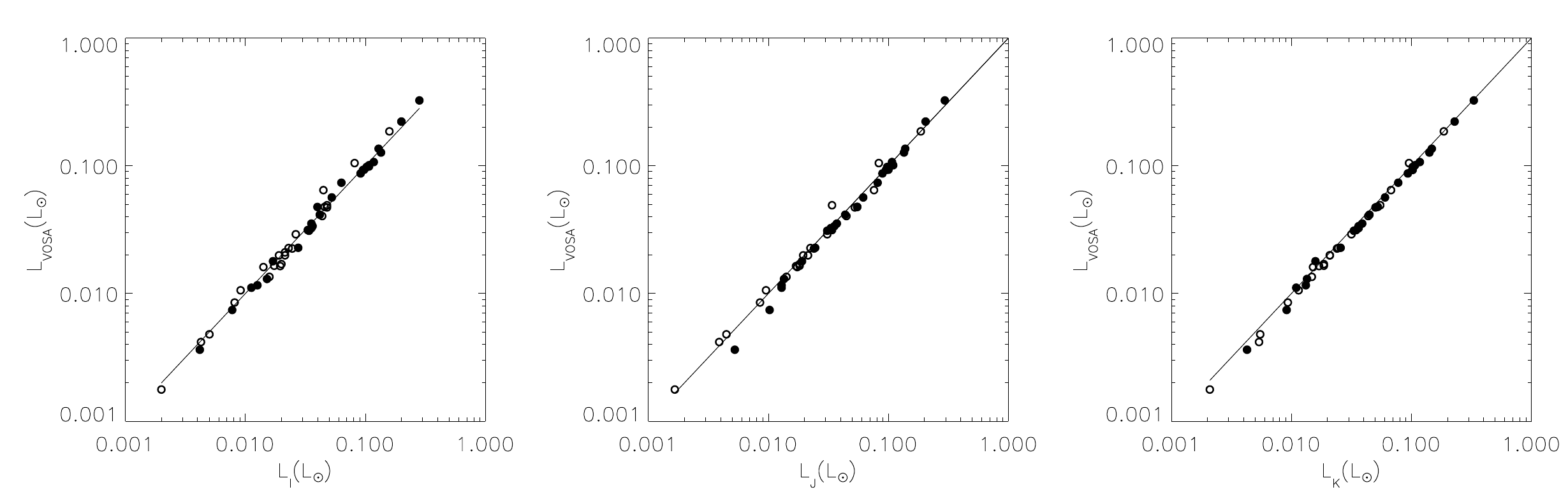}}
\caption{Comparison between the bolometric luminosities estimated with different bolometric corrections applied to various photometrical bands and those obtained with VOSA. Open circles highlight sources harbouring disks.}
\label{llum_comp}
\end{figure}
}

\subsection{Individual sources in C69}
\label{IResults}

As mentioned previously, the location in a T$_{\rm eff}$ vs L$_{\rm bol}$ diagram of 20 of our sources made it impossible for VOSA to estimate their ages and masses. Besides, the age estimated with our methodology for three other sources was much higher than the average age estimated in the literature for Collinder 69. We have divided these 23 ``peculiar sources" in four groups according to the cause that might explain their location in the HR diagram (we have used the same number code as in Table~\ref{all_param}).

\begin{figure}
\resizebox{\hsize}{!}{\includegraphics{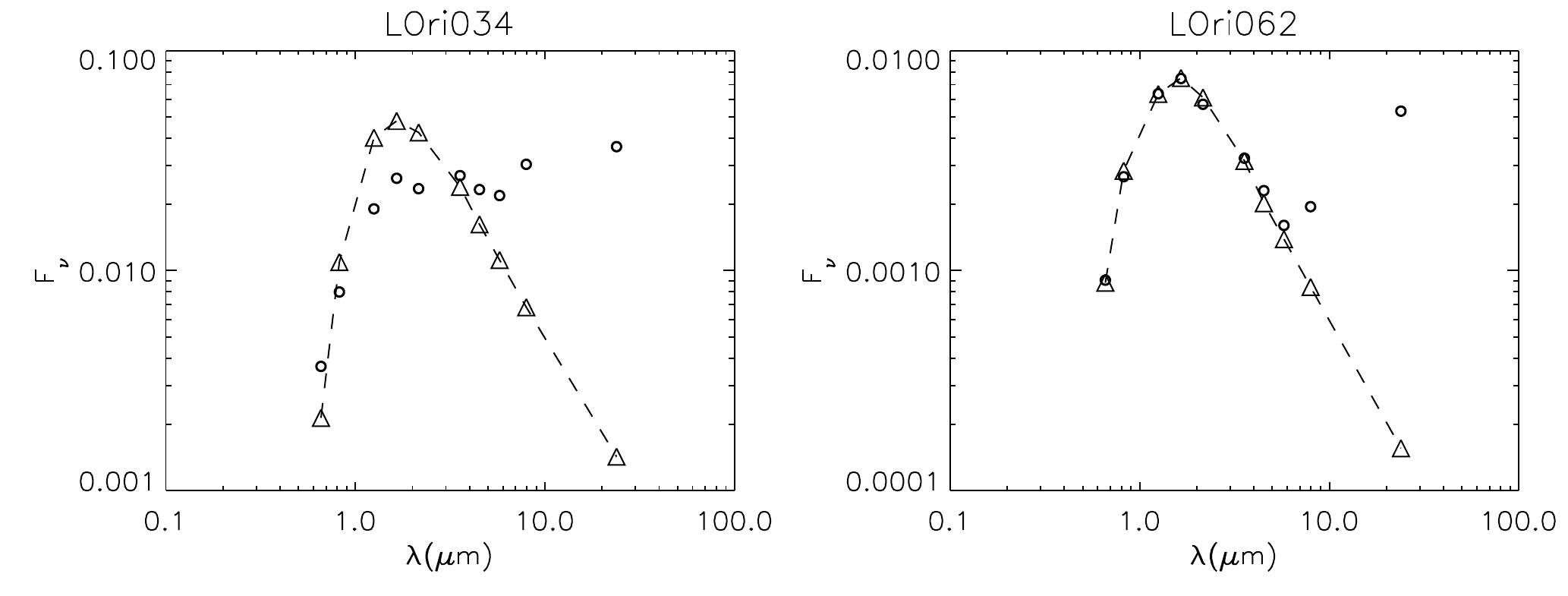}}
\caption{Comparison of the best fits achieved for two sources showing infrared excess (open circles for the observed photometry and open triangles joint with dashed line for the synthetic set). On the left hand side, LOri034, a source whose estimated location in the HR diagram does not allow VOSA to infer neither its mass or its age. On the right hand side, LOri062, another source harbouring a disk but for which al the parameters can be estimated by VOSA and with an inferred age of 12.3 Myr, consistent with the range of ages estimated previously, in the literature, for C69.}
\label{com_disks}
\end{figure}

\begin{figure}
\resizebox{\hsize}{!}{\includegraphics{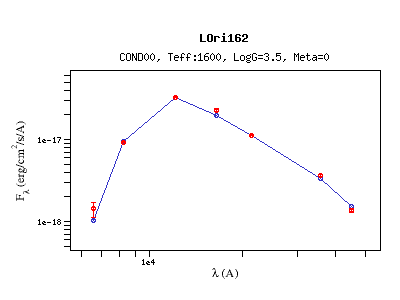}}
\caption{Best fit achieved for LOri162, a possible L field dwarf. We must note that (as mentioned previously in Section~\ref{VOSA}) the SED fitting process is quite insensitive to the surface gravity of the source; besides, at the beginning of the process, we constrained the values of $\log g$ to be between 3.5 and 4.5. Thus, the value of $\log g$ 3.5 shown in the caption should not be taken into account, instead, as reference, the radius assumed when estimating the distance to this source (61-69 parsecs) corresponds to a $\log g$ range of 5.2-5.5}
\label{LOri162}
\end{figure}

\begin{enumerate}
\item Sources with infrared excess caused by the presence of a disk. The dust belonging to the disk, reprocesses the light emitted by the central object, and in the cases where the angle of the disk and/or the amount of flaring is high enough, the light from the star can be blocked (see examples of very low mass stars and brown dwarfs surrounded  by edge-on disks and the effects on the light from the central object in \citealt{Fernandez01,Barrado04a,Pontoppidan05,Luhman07}, and Huelamo et al. 2008, submitted). 

Whenever this is the case, our methodology will underestimate both, the effective temperature and the bolometric luminosity of the central object. We propose this hypothesis as the cause for the peculiar location in the HR diagram of LOri029, LOri034, LOri146, LOri147 and LOri156. 

We must note that we do not see this effect on any of the remaining 24 sources that show infrared excess. See as an example, in Fig~\ref{com_disks}, the comparison between the best fit obtained for one of the sources classified as ``peculiar", LOri034, and the one achieved for LOri062; in the latter case, all the derived parameters, mainly the age, are consistent with the hypothesis that the source belongs to C69. Thus we consider that a 18\% of sources harbouring disks with inclinations or disk structures that block part of the stellar light is a reasonable percentage.

\item Sources previously flagged as possible non-members.
\citet{Barrado07} studied the photometrical properties of the sample, and flagged 16 sources as probable non-members according to these properties. With our methodology, we independently confirm this classification for all 16 sources (see Table~\ref{all_param} for the corresponding IDs); 13 of the cases were very clear, since the location of those sources in the HR diagram fell outside the parameter space confined by the isochrones and evolutionary tracks, in the other three cases (LOri101, LOri141 and LOri165), the derived ages (32, 37 and 85 Myr, respectively) were inconsistent with the upper-limit age derived for the whole cluster (12.3--16 Myr, see Subsection~\ref{C69Results} for details).

\item Degeneracy of the fitting process. In the case of LOri169 no conclusion can be assessed, since the number of points conforming the SED is as low as five (one of them, the R-band magnitude having a very large error), whilst we try to fit four independent parameters.

\item New non-member candidate. Even thought the best fit achieved for LOri162 seems to reproduce quite well the observational data (see Fig~\ref{LOri162}), the estimated luminosity is not compatible at all with the hypothesis that this object belongs to C69. Its luminosity is much higher than the one expected for a cluster member with the derived effective temperature of 1600 K. This luminosity would imply a radius of the star of $\sim$0.5~R$_{\sun}$~(assuming a distance of 400 pc and the relationship: $M_{\rm d}=(R/D)^2$), whilst according to the isochrones from \citet{Baraffe03}, assuming an age of $\sim$10 Myr, the radius of this object should be $\sim$0.15~R$_{\sun}$.

The effective temperature of LOri162 suggests that it might be a field L dwarf that, due to its very low temperature, mimics the colours of the less massive members of Collinder 69. Using the $M_{\rm d}$ factor and assuming a typical radius for 1-5 Gyr dwarfs with the given temperature of 1600 K (according to isochrones from \citealt{Baraffe03}), we estimate a distance to this object between 61 and 69 parsecs and a spectral type of L6-L7 according to the T$_{\rm eff}$ scale by \citet{Basri00}.
\end{enumerate}

\subsection{The cluster C69 as a whole.}
\label{C69Results}

\citet{DM99} studied the $\lambda$ Orionis Star Forming Region (which includes the central cluster Collinder 69 and several younger, dark clouds), and, based on the presence of the Li I $\lambda$6708 absorption (in mid-resolution optical spectra) and different theoretical isochrone sets, they concluded that the star formation in this region (high- and low-mass) began 5--7 Myr ago 
with a dramatic change in the low-mass star formation happening 1--2 Myr ago: either an abrupt cease (through removal of the gas supply), or a burst (caused by star births via cloud compression). 

Later on, \citet{DM01} derived a turnoff age of 6--7 Myr based on Str\"omgren photometry for the massive stars of the cluster. 

Finally, \citet{Barrado07} studied in detail Collinder 69, the older central cluster, and proposed an upper-limit for the age of 20 Myr based on colour-magnitude (optical/near-infrared) diagrams and isochrones by \citet{Baraffe98}, with an optimal age of 5 Myr for the isochrones \citep{Baraffe02} corresponding to the DUSTY models.

The methodology presented in this work, provides us with an estimation of the age of the association through the analysis of the statistics of the ages inferred for each individual member of the cluster. We have selected a subsample of 120 sources (out of the total of 167 candidates) that conforms the ``least-possible biased" set in the sense that we do not include either sources flagged as possible non-member or sources with infrared excesses whose age estimation might be affected by the effect of the disk on the apparent luminosity of the central object. This set of sources can be considered as a representative sample of the Class III population of Collinder 69 in the 0.014--1.15M$_\odot$ range of masses. From the 120 sources, the SEDs of three of them are best reproduced with models from the DUSTY collection, while for the remaining 117 the best fitting model always corresponds to either the NextGen or the  Kurucz collection. Since different sets of isochrones and evolutionary tracks have to be applied when using the DUSTY models or the NextGen and Kurucz models, we will split our sample in two sets: 
\begin{itemize}
\item For the three sources where the DUSTY models reproduce the best the observed SEDs, we derive an upper-limit age of 5 Myr
\item Among the remaining 117 sources, we find that while 90\% of them are younger than 16 Myr, the 3rd Quartile is located at 12.3 Myr; therefore we infer an upper-limit age for Collinder 69 of 12.3-16 Myr.
\end{itemize}


This final upper-limit  of 12.3-16 Myr is consistent with the estimated ages for the cluster by other authors.

A limitation of our methodology is that, since we are performing fitting with four free parameters, a minimum number of six data-points in the observed SED is desirable (to avoid degeneracy). Another 
issue to take into account is that
 our results (mainly the derivation of the effective temperatures) are quite discrete, since the minimum step in temperature in the grid of models is 100K. 
 
 On the other hand, a great advantage is that the process followed is completely automatic; therefore, if a new (denser) grid of models is published, new observational data become available or new candidate members are detected, our determination of the age can be easily and quickly re-computed.

\section{Conclusions}
\label{C}

We have presented a new VO-tool, VOSA, developed by the Spanish Virtual Observatory team for this particular scientific case (but with a wider applicability range) to easy the process of calculating synthetic photometry and performing $\chi^{2}$ tests to large sets of fittings as well as to interpolate among collections of isochrones and evolutionary tracks.

When applied to the case of Collinder 69, we have been able to perform the following studies:
\begin{enumerate}
\item We have estimated different physical parameters (effective temperatures, gravity, bolometric luminosity, and in most of the cases, mass and age) for 167 candidate members.
\item We have independently confirmed the classification from \citet{Barrado07} of non-members for 16 of the sources of the sample, and we have added a new possible non-member to this list (LOri162).
\item We have derived an upper-limit for the age of 12.3-16 Myr consistent with previous estimations in the literature.
\end{enumerate}



We must note that VOSA is a very efficient tool; as an example, it only takes $\sim$ 20 minutes to reproduce the whole workflow presented in this paper.


\begin{acknowledgements}
This research has been funded by Spanish grants MEC/ESP 2007-65475-C02-02, MEC/Consolider-CSD2006-0070,
and CAM/PRICIT-S-0505/ESP/0361, and has made use of the Spanish Virtual Observatory supported from 
the Spanish MEC through grants AyA2005-04286, AyA2005-24102-E.
A. Bayo wishes to acknowledge the Spanish Ministry of Science and Innovation for the financial 
support of a graduate fellowship.
\end{acknowledgements}


\bibliographystyle{aa}
\bibliography{biblio}

\section*{Appendix A: Synthetic photometry.}

\begin{itemize}
\item Starting point: observed magnitudes (in filters $G_i$) that are translated with the zero point $ZP$ into Jy 
. Thus, we have a flux density per unit of frequency, that is $F_{\nu,G_i}$. We can also use $ZP$ for $F_{\lambda}$ and then have another flux density
(in this case $F_{\lambda,G_i}$).

%

\smallskip

\item The models are functions $F(\lambda)$ with values in $\rm{erg/cm^2/s/\AA}$, the filters are functions $G_1(\lambda)$ without physical units, but when we normalize the response curves:\\
 
\smallskip

$N_1(\lambda)=G_1(\lambda)/\int_\lambda G_1(\lambda') d\lambda'$, \\

\smallskip

this new function $N_1(\lambda)$ is not dimensionless anymore, it has physical units of $\lambda^{-1}$. This is the function that we multiply by the model to obtain a synthetic flux density. So,\\

\smallskip

$F_{\rm{syn,N}}=\int_\lambda F_{\rm{mod},\lambda}(\lambda) \times N_1(\lambda) d\lambda$,\\

\smallskip

with units of $\rm{erg/cm^2/s/\AA}$.

\smallskip

\item Our observational data-point corresponds to a {\bf weighted} average of the flux observed through a filter. The process is:

\smallskip

\begin{itemize}
\item The light ($F_*$) passes through the filter (a function with values between 0 and 1): $F_*\times G_1(\lambda)$

\smallskip

\item This ``total amount of light" is translated into a density by dividing by the area enclosed by the filter: $\int_\lambda G_1(\lambda) d\lambda$

\end{itemize}

\smallskip

\item To recover the total flux that was observed through the filter $G_1$, we multiply the ``observed flux density" by the integral of the flux:\\

 \smallskip

$F_{\rm{total,G_1}}=F_{G_1}\times \int_\lambda G_1(\lambda) d\lambda$. \\

\smallskip

This $F_{\rm{total,G_1}}$ has the required physical units of a flux: $\rm{erg/cm^2/s}$.

\smallskip

\item If we want to compute the bolometric correction using the model, we integrate it: \\

\smallskip

$F_M=\int_\lambda F(\lambda) d\lambda$,\\

\smallskip

 subtract the density fluxes corresponding to each filter multiplied by the areas of the filters: \\

 \smallskip

$F_M - \sum_i (F_{syn,N_i}\times \int_\lambda G_i(\lambda) d\lambda)$),\\

\smallskip

 and add the $F_{\rm{total,G_i}}$ calculated in the previous step. \\

 This takes into the account the model to correct from possible intersections among the filters wavelength coverage.
 
\end{itemize}

\section*{Appendix B: Workflow.}

\begin{figure}
\resizebox{\hsize}{!}{\includegraphics{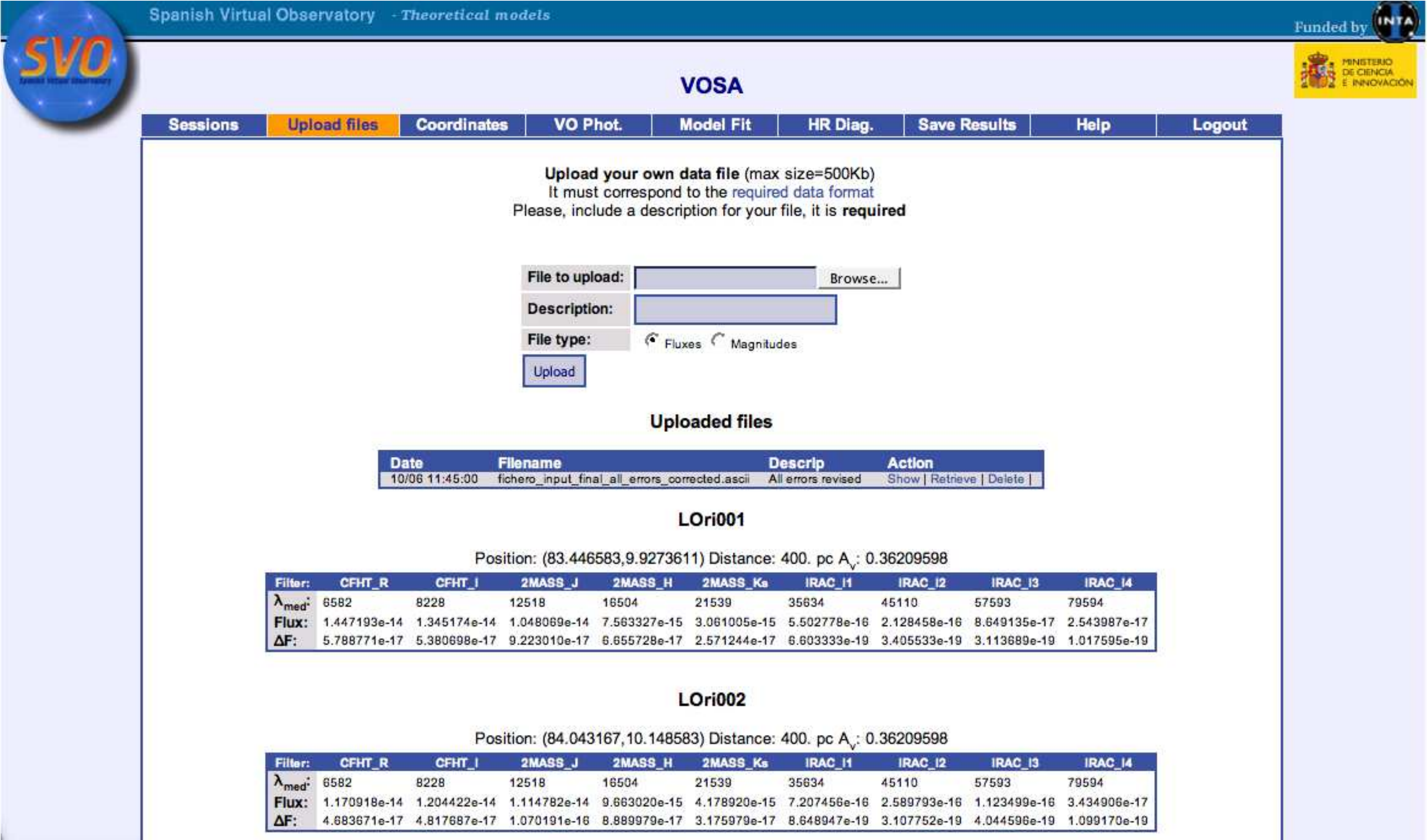}}
\caption{Part of the input data in the web interface of the SVO SED fitting tool.}
\label{fig:input1}
\end{figure}

\begin{figure}
\resizebox{\hsize}{!}{\includegraphics{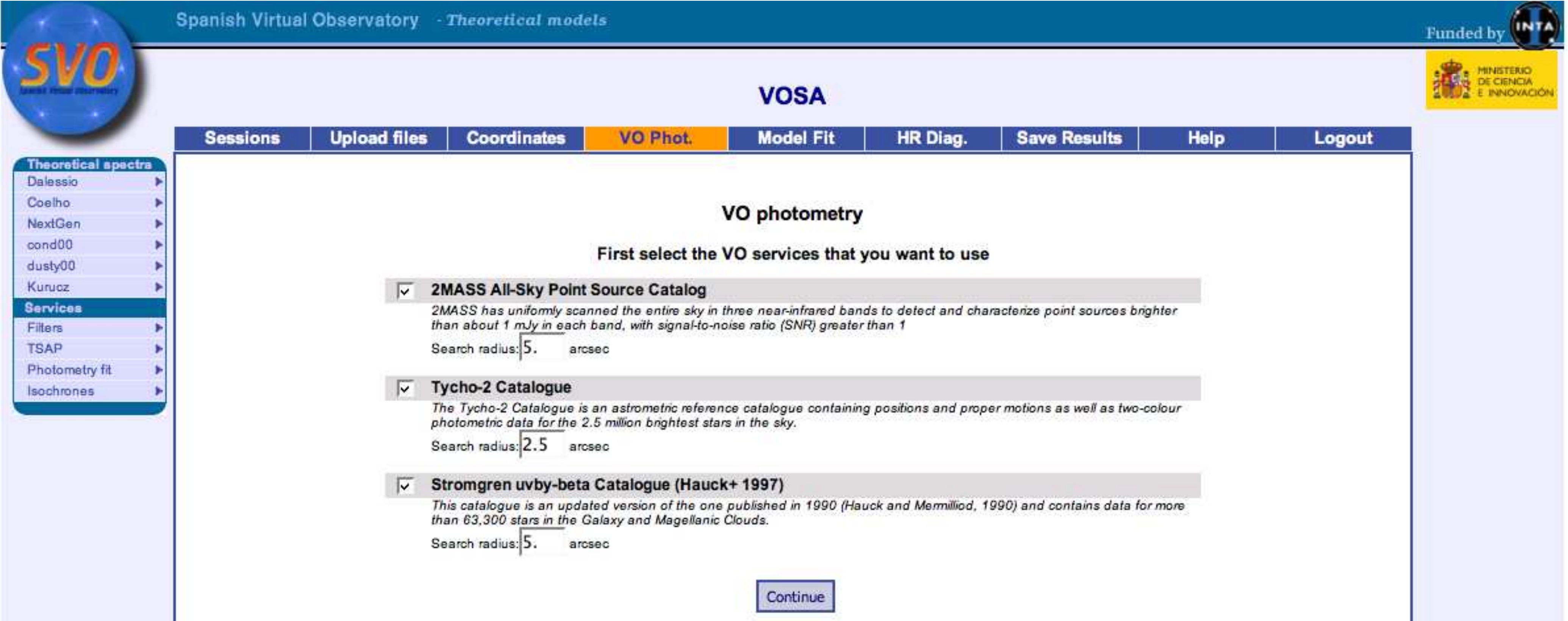}}
\caption{Photometrical catalogs to be queried with the respective search radii.}
\label{VOphot}
\end{figure}

\begin{figure}
\resizebox{\hsize}{!}{\includegraphics{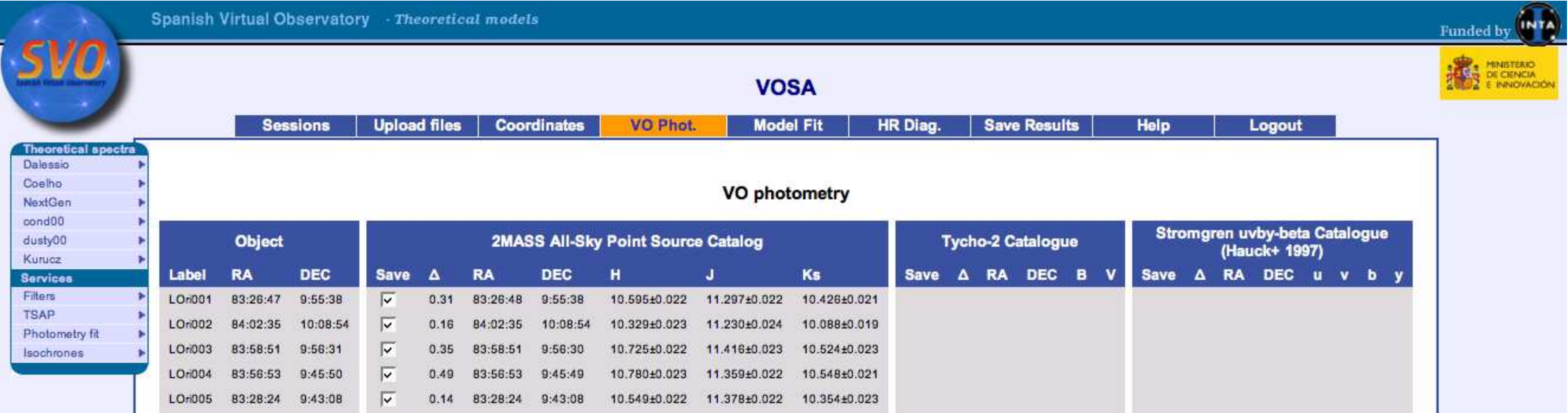}}
\caption{2MASS photometry found. We will use these JHKs measurements in those objects for which we do not have our own photometrical data.}
\label{VOphot2}
\end{figure}

{\scriptsize
\longtab{2}{ 
\tabcolsep 4.5pt
\begin{longtable}{llcrrrrrrr}
\caption{\label{all_param} Parameters derived for each individual source of the sample. For some of the sources, we have not been able to interpolate between the isochrones and evolutionary tracks (to infer ages and masses), because of their peculiar location in the HR diagram; we have grouped these sources in four possible categories according to the probable cause for this peculiar location (last column of the table): 
[1] Source with disk, 
[2] Previous ``non-member" candidates according to the photometric properties (see \citealt{Barrado07} for further details), 
[3] Degenerated fit due the small number of points to be fitted vs number of free parameters, and 
[4] possible foreground cool source 
(see Section ~\ref{IResults} for details on each individual case).
[*] For a more detailed analysis in LOri167 see \citet{BarradoL07}}\\
\hline\hline 
{ID} &{Model} &{T$_{\rm eff}$} &{logg} &{F$_{\rm obs}$/F$_{Tot}$} &{L(L$_\odot$)}&{$\Delta$L(L$_\odot$)} &{Age(Myr)} &{M(M$_\odot$)}&Note \\
\hline 
\endfirsthead 
\caption{continued.}\\ 
\hline\hline 
{ID} &{Model} &{T$_{\rm eff}$} &{logg} &{F$_{\rm obs}$/F$_{Tot}$} &{L(L$_\odot$)}&{$\Delta$L(L$_\odot$)} &{Age(Myr)} &{M(M$_\odot$)} &Note\\
\hline 
\endhead 
\hline 
\endfoot 
  LOri001 & Kurucz   & 4000 & 3.5 & 0.48 &   9.52e-1   &2.80e-3 & 3.0 & 1.1265& -\\
  LOri002 & NextGen  & 3800 & 4.5 & 0.48 &   1.02e+0   &3.18e-3 & 1.6 & 0.9508& -\\
  LOri003 & Kurucz   & 4000 & 3.5 & 0.47 &   8.78e-1   &2.55e-3 & 3.2 & 1.105& -\\
  LOri004 & Kurucz   & 3750 & 4.0 & 0.48 &   8.23e-1   &2.45e-3 & 1.7 & 0.8613& -\\
  LOri005 & NextGen  & 4000 & 4.0 & 0.48 &   9.10e-1   &2.73e-3 & 3.2 & 1.113& -\\
  LOri006 & Kurucz   & 4000 & 3.5 & 0.47 &   7.66e-1   &2.45e-3 & 4.0 & 1.0983& -\\
  LOri007 & Kurucz   & 4000 & 4.5 & 0.48 &   6.61e-1   &2.17e-3 & 5.0 & 1.0471& -\\
  LOri008 & Kurucz   & 4000 & 3.5 & 0.48 &   7.44e-1   &2.45e-3 & 4.0 & 1.0962& -\\
  LOri009 & NextGen  & 4000 & 3.5 & 0.49 &   5.96e-1   &1.88e-3 & 5.8 & 1.0505& -\\
  LOri010 & NextGen  & 4200 & 4.0 & 0.48 &   5.73e-1   &1.84e-3 & 10.0 & 1.1466& -\\
  LOri011 & NextGen  & 3900 & 4.5 & 0.47 &   7.06e-1   &2.33e-3 & 3.7 & 1.037& -\\
  LOri012 & NextGen  & 4000 & 4.5 & 0.48 &   6.22e-1   &2.00e-3 & 5.2 & 1.0544& -\\
  LOri013 & NextGen  & 3700 & 4.5 & 0.49 &   6.12e-1   &1.93e-3 & 2.3 & 0.7737& -\\
  LOri014 & Kurucz   & 4000 & 4.5 & 0.49 &   5.37e-1   &1.74e-3 & 6.4 & 1.0473& -\\
  LOri015 & Kurucz   & 4000 & 3.5 & 0.49 &   5.60e-1   &1.77e-3 & 6.3 & 1.0495& -\\
  LOri016 & Kurucz   & 3750 & 4.5 & 0.47 &   5.06e-1   &1.65e-3 & 3.4 & 0.8185& -\\
  LOri017 & NextGen  & 4200 & 4.0 & 0.48 &   4.52e-1   &1.36e-3 & 13.7 & 1.0499& -\\
  LOri018 & Kurucz   & 3750 & 3.5 & 0.47 &   4.91e-1   &1.51e-3 & 3.7 & 0.8153& -\\
  LOri019 & Kurucz   & 3750 & 3.5 & 0.48 &   4.65e-1   &1.52e-3 & 4.0 & 0.8102& -\\
  LOri020 & NextGen  & 3600 & 4.5 & 0.48 &   4.86e-1   &1.75e-3 & 2.0 & 0.624& -\\
  LOri021 & Kurucz   & 4000 & 3.5 & 0.47 &   4.10e-1   &1.33e-3 & 9.9 & 0.9998& -\\
  LOri022 & Kurucz   & 3750 & 4.0 & 0.48 &   4.26e-1   &1.29e-3 & 4.5 & 0.8043& -\\
  LOri023 & NextGen  & 4000 & 4.5 & 0.47 &   4.08e-1   &1.30e-3 & 10.0 & 0.9996& -\\
  LOri024 & Kurucz   & 3750 & 3.5 & 0.48 &   4.11e-1   &1.50e-3 & 4.9 & 0.8024& -\\
  LOri025 & NextGen  & 3700 & 4.5 & 0.44 &   4.71e-1   &2.25e-3 & 3.2 & 0.7504& -\\
  LOri026 & NextGen  & 3700 & 4.5 & 0.48 &   4.30e-1   &1.51e-3 & 3.7 & 0.75& -\\
  LOri027 & NextGen  & 4000 & 4.5 & 0.48 &   3.51e-1   &1.11e-3 & 12.5 & 0.9586& -\\
  LOri028 & Kurucz   & 3750 & 4.0 & 0.49 &   2.93e-1   &9.22e-4 & 7.4 & 0.7584& -\\
  LOri029 & NextGen  & 3100 & 4.5 & 0.4  &   4.49e-1   &1.27e-3 &  & & [1]\\
  LOri030 & NextGen  & 3700 & 4.5 & 0.48 &   3.19e-1   &1.11e-3 & 5.3 & 0.7237& -\\
  LOri031 & NextGen  & 3800 & 4.5 & 0.47 &   3.25e-1   &1.12e-3 & 7.7 & 0.8366& -\\
  LOri032 & NextGen  & 3700 & 4.5 & 0.48 &   3.04e-1   &1.05e-3 & 5.9 & 0.7128& -\\
  LOri033 & NextGen  & 3700 & 4.5 & 0.46 &   3.17e-1   &1.34e-3 & 5.3 & 0.7231& -\\
  LOri034 & NextGen  & 3000 & 4.0 & 0.36 &   4.32e-1   &1.08e-3 &  & & [1]\\
  LOri035 & NextGen  & 3700 & 4.5 & 0.48 &   2.67e-1   &9.09e-4 & 6.6 & 0.7005& -\\
  LOri036 & NextGen  & 3600 & 4.5 & 0.49 &   2.49e-1   &8.21e-4 & 5.0 & 0.5883& -\\
  LOri037 & NextGen  & 3700 & 4.5 & 0.48 &   2.82e-1   &9.78e-4 & 6.3 & 0.7033& -\\
  LOri038 & NextGen  & 3800 & 4.5 & 0.4  &   2.55e-1   &7.92e-4 & 10.1 & 0.8002& -\\
  LOri039 & NextGen  & 3800 & 4.5 & 0.47 &   2.41e-1   &8.26e-4 & 11.1 & 0.7997& -\\
  LOri040 & NextGen  & 3600 & 4.5 & 0.48 &   2.56e-1   &8.55e-4 & 5.0 & 0.5922& -\\
  LOri041 & NextGen  & 3400 & 4.5 & 0.49 &   2.54e-1   &8.89e-4 & 2.0 & 0.3932& -\\
  LOri042 & NextGen  & 3800 & 4.5 & 0.47 &   2.22e-1   &7.69e-4 & 12.6 & 0.7929& -\\
  LOri043 & NextGen  & 3600 & 4.5 & 0.49 &   2.25e-1   &7.95e-4 & 5.6 & 0.5744& -\\
  LOri044 & Kurucz   & 3750 & 3.5 & 0.48 &   1.94e-1   &6.37e-4 & 12.6 & 0.7261& -\\
  LOri045 & NextGen  & 3600 & 4.5 & 0.49 &   2.07e-1   &7.30e-4 & 6.3 & 0.5703& -\\
  LOri046 & NextGen  & 3800 & 4.5 & 0.47 &   1.71e-1   &5.72e-4 & 18.0 & 0.7508& -\\
  LOri047 & NextGen  & 3400 & 4.5 & 0.47 &   2.03e-1   &7.80e-4 & 2.5 & 0.3614& -\\
  LOri048 & NextGen  & 3500 & 4.5 & 0.47 &   1.90e-1   &7.20e-4 & 4.3 & 0.4516& -\\
  LOri049 & NextGen  & 3800 & 4.5 & 0.47 &   1.52e-1   &5.52e-4 & 20.4 & 0.7471& -\\
  LOri050 & NextGen  & 3300 & 4.5 & 0.44 &   1.86e-1   &6.76e-4 & 1.9 & 0.2996& -\\
  LOri051 & NextGen  & 3700 & 4.5 & 0.48 &   1.44e-1   &4.64e-4 & 15.8 & 0.6508& -\\
  LOri052 & NextGen  & 3600 & 4.5 & 0.48 &   1.52e-1   &5.03e-4 & 9.5 & 0.5604& -\\
  LOri053 & NextGen  & 3600 & 4.5 & 0.48 &   1.40e-1   &5.42e-4 & 10.0 & 0.5588& -\\
  LOri054 & NextGen  & 3500 & 4.5 & 0.48 &   1.38e-1   &4.66e-4 & 6.5 & 0.45& -\\
  LOri055 & NextGen  & 3600 & 4.5 & 0.49 &   1.36e-1   &5.28e-4 & 10.4 & 0.5584& -\\
  LOri056 & NextGen  & 3400 & 4.5 & 0.47 &   1.27e-1   &4.93e-4 & 4.6 & 0.35& -\\
  LOri057 & NextGen  & 3400 & 4.5 & 0.46 &   1.07e-1   &3.76e-4 & 5.9 & 0.35& -\\
  LOri058 & NextGen  & 3500 & 4.5 & 0.49 &   9.87e-2   &3.55e-4 & 10.5 & 0.4498& -\\
  LOri059 & NextGen  & 3500 & 4.5 & 0.48 &   9.79e-2   &3.84e-4 & 10.7 & 0.4498& -\\
  LOri060 & NextGen  & 3500 & 4.5 & 0.49 &   9.28e-2   &4.03e-4 & 12.2 & 0.4498& -\\
  LOri061 & NextGen  & 3300 & 4.5 & 0.44 &   1.06e-1   &3.54e-4 & 3.7 & 0.2753& -\\
  LOri062 & NextGen  & 3500 & 4.5 & 0.46 &   9.20e-2   &3.54e-4 & 12.3 & 0.4498& -\\
  LOri063 & NextGen  & 3100 & 4.0 & 0.39 &   1.05e-1   &3.37e-4 & 1.5 & 0.157& -\\
  LOri064 & NextGen  & 3500 & 4.5 & 0.48 &   8.10e-2   &3.05e-4 & 14.9 & 0.4496& -\\
  LOri065 & NextGen  & 3400 & 4.5 & 0.49 &   7.76e-2   &3.10e-4 & 9.6 & 0.3503& -\\
  LOri066 & NextGen  & 3200 & 4.5 & 0.5  &   8.87e-2   &3.66e-4 & 2.9 & 0.2098& -\\
  LOri067 & NextGen  & 3500 & 4.5 & 0.49 &   6.44e-2   &2.94e-4 & 19.9 & 0.4469& -\\
  LOri068 & NextGen  & 3400 & 4.5 & 0.47 &   9.32e-2   &3.52e-4 & 7.1 & 0.3501& -\\
  LOri069 & NextGen  & 3200 & 4.5 & 0.5  &   1.04e-1   &4.29e-4 & 2.4 & 0.2151& -\\
  LOri070 & NextGen  & 3400 & 4.5 & 0.49 &   6.13e-2   &2.78e-4 & 12.7 & 0.35& -\\
  LOri071 & NextGen  & 3200 & 4.5 & 0.48 &   7.37e-2   &2.98e-4 & 3.4 & 0.2028& -\\
  LOri072 & NextGen  & 3100 & 4.5 & 0.43 &   9.69e-2   &3.61e-4 & 1.6 & 0.1614& -\\
  LOri073 & NextGen  & 3400 & 4.5 & 0.46 &   8.71e-2   &3.15e-4 & 7.9 & 0.3502& -\\
  LOri074 & NextGen  & 3300 & 4.5 & 0.49 &   8.33e-2   &3.18e-4 & 4.9 & 0.2619& -\\
  LOri075 & NextGen  & 3200 & 4.5 & 0.46 &   1.01e-1   &3.72e-4 & 2.5 & 0.2144& -\\
  LOri076 & NextGen  & 3400 & 4.5 & 0.48 &   5.31e-2   &2.13e-4 & 15.8 & 0.3485& -\\
  LOri077 & NextGen  & 3200 & 4.5 & 0.5  &   5.65e-2   &2.57e-4 & 4.3 & 0.1965& -\\
  LOri078 & NextGen  & 3300 & 4.5 & 0.48 &   5.28e-2   &3.26e-4 & 8.1 & 0.2502& -\\
  LOri079 & NextGen  & 3300 & 4.5 & 0.5  &   4.84e-2   &2.43e-4 & 9.5 & 0.25& -\\
  LOri080 & NextGen  & 3100 & 4.0 & 0.49 &   6.45e-2   &2.56e-4 & 2.5 & 0.1519& -\\
  LOri081 & NextGen  & 3200 & 4.5 & 0.42 &   4.91e-2   &1.96e-4 & 5.0 & 0.1783& -\\
  LOri082 & NextGen  & 3300 & 4.5 & 0.5  &   4.77e-2   &2.34e-4 & 9.7 & 0.25& -\\
  LOri083 & NextGen  & 3300 & 4.5 & 0.49 &   4.69e-2   &2.41e-4 & 9.8 & 0.25& -\\
  LOri084 & NextGen  & 3300 & 4.5 & 0.51 &   5.18e-2   &2.36e-4 & 8.3 & 0.2501& -\\
  LOri085 & NextGen  & 3100 & 4.5 & 0.43 &   5.32e-2   &2.12e-4 & 3.1 & 0.1497& -\\
  LOri086 & NextGen  & 3400 & 4.5 & 0.49 &   4.06e-2   &2.04e-4 & 20.1 & 0.315& -\\
  LOri087 & NextGen  & 3300 & 4.5 & 0.5  &   4.76e-2   &2.49e-4 & 9.7 & 0.25& -\\
  LOri088 & NextGen  & 3200 & 4.5 & 0.51 &   4.80e-2   &2.64e-4 & 5.0 & 0.1762& -\\
  LOri089 & NextGen  & 3300 & 4.5 & 0.5  &   4.04e-2   &2.15e-4 & 11.7 & 0.2482& -\\
  LOri090 & NextGen  & 3300 & 4.5 & 0.46 &   3.89e-2   &1.83e-4 & 12.4 & 0.2471& -\\
  LOri091 & NextGen  & 3100 & 4.0 & 0.49 &   4.72e-2   &2.32e-4 & 3.2 & 0.1432& -\\
  LOri092 & NextGen  & 3300 & 4.5 & 0.47 &   3.94e-2   &1.92e-4 & 12.2 & 0.2475& -\\
  LOri093 & NextGen  & 3300 & 4.5 & 0.49 &   3.91e-2   &2.24e-4 & 12.3 & 0.2473& -\\
  LOri094 & NextGen  & 3200 & 4.0 & 0.48 &   4.14e-2   &2.10e-4 & 5.8 & 0.1743& -\\
  LOri095 & NextGen  & 3300 & 4.5 & 0.5  &   3.53e-2   &1.95e-4 & 12.8 & 0.243& -\\
  LOri096 & NextGen  & 3100 & 4.5 & 0.44 &   3.66e-2   &1.89e-4 & 4.0 & 0.1295& -\\
  LOri098 & NextGen  & 3400 & 4.0 & 0.49 &   3.37e-2   &2.09e-4 & 25.0 & 0.3& -\\
  LOri099 & NextGen  & 3400 & 4.0 & 0.47 &   3.25e-2   &1.64e-4 & 25.1 & 0.2998& -\\
  LOri100 & NextGen  & 3300 & 4.5 & 0.46 &   3.10e-2   &1.80e-4 & 15.7 & 0.2337& -\\
  LOri101 & NextGen  & 3400 & 4.5 & 0.45 &   2.60e-2   &1.49e-4 & 31.7 & 0.2989& -\\
  LOri102 & NextGen  & 3200 & 4.0 & 0.48 &   3.27e-2   &2.34e-4 & 7.6 & 0.1653& -\\
  LOri103 & NextGen  & 3200 & 4.5 & 0.49 &   2.99e-2   &1.50e-4 & 8.0 & 0.162& -\\
  LOri104 & NextGen  & 3000 & 4.0 & 0.45 &   3.10e-2   &1.54e-4 & 3.1 & 0.0899& -\\
  LOri105 & NextGen  & 3200 & 4.0 & 0.49 &   2.51e-2   &1.82e-4 & 10.0 & 0.1565& -\\
  LOri106 & NextGen  & 3100 & 4.0 & 0.45 &   2.91e-2   &1.92e-4 & 4.9 & 0.116& -\\
  LOri107 & NextGen  & 3000 & 3.5 & 0.49 &   3.14e-2   &2.08e-4 & 3.1 & 0.0899& -\\
  LOri108 & NextGen  & 3100 & 4.5 & 0.5  &   2.56e-2   &1.75e-4 & 5.1 & 0.1103& -\\
  LOri109 & NextGen  & 3200 & 4.5 & 0.52 &   2.28e-2   &5.08e-5 & 11.2 & 0.1544& -\\
  LOri110 & NextGen  & 3400 & 4.0 & 0.48 &   2.25e-2   &1.74e-4 & 39.8 & 0.2947& -\\
  LOri111 & NextGen  & 3000 & 4.0 & 0.5  &   2.66e-2   &1.86e-4 & 3.4 & 0.0882& -\\
  LOri112 & NextGen  & 3100 & 4.0 & 0.48 &   2.49e-2   &1.77e-4 & 5.2 & 0.1101& -\\
  LOri113 & NextGen  & 3200 & 4.5 & 0.49 &   1.99e-2   &4.14e-5 & 12.7 & 0.1513& -\\
  LOri114 & NextGen  & 3000 & 4.0 & 0.48 &   2.27e-2   &1.73e-4 & 4.0 & 0.0822& -\\
  LOri115 & NextGen  & 3300 & 4.5 & 0.48 &   1.64e-2   &1.48e-4 & 31.4 & 0.2& -\\
  LOri116 & NextGen  & 3200 & 4.0 & 0.47 &   1.75e-2   &1.46e-4 & 15.6 & 0.1501& -\\
  LOri117 & NextGen  & 3100 & 3.5 & 0.5  &   2.10e-2   &4.51e-5 & 6.3 & 0.1096& -\\
  LOri118 & NextGen  & 3000 & 4.0 & 0.43 &   1.99e-2   &1.36e-4 & 4.3 & 0.08& -\\
  LOri119 & NextGen  & 3100 & 4.0 & 0.54 &   1.79e-2   &5.97e-5 & 7.8 & 0.1084& -\\
  LOri120 & NextGen  & 3100 & 4.0 & 0.46 &   1.65e-2   &1.38e-4 & 8.1 & 0.1072& -\\
  LOri121 & NextGen  & 3200 & 4.5 & 0.48 &   1.34e-2   &1.35e-4 & 20.0 & 0.1476& -\\
  LOri122 & NextGen  & 3100 & 4.0 & 0.46 &   1.52e-2   &1.37e-4 & 9.5 & 0.1057& -\\
  LOri124 & NextGen  & 3200 & 4.0 & 0.45 &   1.30e-2   &1.36e-4 & 20.4 & 0.1464& -\\
  LOri125 & NextGen  & 3200 & 4.0 & 0.46 &   1.26e-2   &1.40e-4 & 21.3 & 0.1452& -\\
  LOri126 & NextGen  & 3000 & 4.5 & 0.5  &   1.35e-2   &2.80e-5 & 6.2 & 0.0736& -\\
  LOri127 & DUSTY00  & 1900 & 4.5 & 0.74 &   6.04e-2   &4.21e-4 &  & & [2]\\
  LOri128 & NextGen  & 3100 & 4.0 & 0.49 &   1.26e-2   &1.67e-4 & 11.9 & 0.1022& -\\
  LOri129 & NextGen  & 3000 & 3.5 & 0.44 &   1.61e-2   &1.48e-4 & 5.0 & 0.075& -\\
  LOri130 & NextGen  & 3200 & 4.0 & 0.47 &   1.16e-2   &1.23e-4 & 24.9 & 0.1439& -\\
  LOri131 & NextGen  & 3000 & 3.5 & 0.5  &   1.39e-2   &1.44e-4 & 6.1 & 0.0742& -\\
  LOri132 & NextGen  & 3000 & 3.5 & 0.49 &   1.21e-2   &1.53e-4 & 6.6 & 0.0723& -\\
  LOri133 & NextGen  & 3800 & 3.5 & 0.43 &   7.85e-3   &1.24e-4 &  & & [2]\\
  LOri134 & NextGen  & 3100 & 3.5 & 0.49 &   1.17e-2   &1.32e-4 & 12.6 & 0.1017& -\\
  LOri135 & NextGen  & 3100 & 3.5 & 0.46 &   1.11e-2   &1.38e-4 & 12.9 & 0.1015& -\\
  LOri136 & NextGen  & 2900 & 3.5 & 0.5  &   1.24e-2   &1.80e-4 & 5.0 & 0.0602& -\\
  LOri137 & NextGen  & 4000 & 4.0 & 0.31 &   3.50e-3   &3.94e-5 &  & & [2]\\
  LOri138 & NextGen  & 3000 & 4.0 & 0.49 &   9.78e-3   &1.28e-4 & 8.8 & 0.0738& -\\
  LOri139 & NextGen  & 3000 & 4.0 & 0.49 &   8.50e-3   &1.84e-5 & 10.1 & 0.0745& -\\
  LOri140 & NextGen  & 2800 & 3.5 & 0.43 &   1.06e-2   &1.23e-4 & 4.0 & 0.0405& -\\
  LOri141 & NextGen  & 3100 & 4.0 & 0.5  &   5.69e-3   &1.67e-5 & 36.8 & 0.1088& -\\
  LOri142 & NextGen  & 3000 & 4.0 & 0.51 &   7.33e-3   &1.92e-5 & 12.6 & 0.0749& -\\
  LOri143 & NextGen  & 3000 & 4.0 & 0.53 &   7.43e-3   &2.29e-5 & 12.6 & 0.0749& -\\
  LOri144 & NextGen  & 3800 & 4.5 & 0.54 &   2.85e-3   &4.29e-5 &  & & [2]\\
  LOri146 & NextGen  & 2700 & 3.5 & 0.44 &   8.00e-3   &1.32e-4 &  & & [1]\\
  LOri147 & NextGen  & 3500 & 3.5 & 0.51 &   5.24e-3   &2.00e-5 &  & & [1]\\
  LOri148 & NextGen  & 3000 & 4.5 & 0.54 &   5.19e-3   &1.63e-5 & 20.4 & 0.0756& -\\
  LOri149 & Kurucz   & 3500 & 3.5 & 0.24 &   2.12e-3   &4.22e-6 &  & & [2]\\
  LOri150 & NextGen  & 2800 & 3.5 & 0.48 &   4.79e-3   &1.31e-4 & 9.8 & 0.0472& -\\
  LOri151 & NextGen  & 3200 & 4.5 & 0.51 &   2.81e-3   &1.48e-5 &  & & [2]\\
  LOri152 & NextGen  & 3500 & 3.5 & 0.5  &   3.26e-3   &1.17e-4 &  & & [2]\\
  LOri153 & NextGen  & 2900 & 4.5 & 0.52 &   3.49e-3   &1.38e-5 & 25.1 & 0.0659& -\\
  LOri154 & NextGen  & 2700 & 3.5 & 0.48 &   4.17e-3   &1.20e-4 & 7.9 & 0.0315& -\\
  LOri155 & NextGen  & 2700 & 4.5 & 0.53 &   3.63e-3   &1.11e-5 & 9.4 & 0.0336& -\\
  LOri156 & COND00   & 1500 & 3.5 & 0.54 &   3.20e-3   &1.43e-5 &  & & [1]\\
  LOri157 & NextGen  & 3700 & 3.5 & 0.47 &   1.54e-3   &9.04e-6 &  & & [2]\\
  LOri158 & NextGen  & 3900 & 3.5 & 0.45 &   1.18e-3   &8.72e-6 &  & & [2]\\
  LOri159 & NextGen  & 3000 & 4.5 & 0.49 &   1.28e-3   &1.17e-5 &  & & [2]\\
  LOri160 & NextGen  & 2700 & 3.5 & 0.5  &   1.65e-3   &1.02e-5 & 31.6 & 0.0467& -\\
  LOri161 & NextGen  & 2600 & 4.5 & 0.52 &   1.77e-3   &1.37e-5 & 16.0 & 0.0308& -\\
  LOri162 & COND00   & 1600 & 3.5 & 0.6  &   1.71e-3   &2.37e-5 &  & & [4]\\
  LOri163 & DUSTY00  & 2000 & 4.0 & 0.55 &   1.65e-3   &2.13e-5 & 1.0 & 0.006 & -\\
  LOri164 & NextGen  & 3800 & 3.5 & 0.45 &   7.29e-4   &5.92e-6 &  & & [2]\\
  LOri165 & DUSTY00  & 2500 & 3.5 & 0.51 &   7.50e-4   &1.50e-5 & 85.3 & 0.048& [2]\\
  LOri166 & DUSTY00  & 2200 & 4.5 & 0.43 &   9.35e-4   &6.94e-6 & 4.7 & 0.0136& -\\
  LOri167 & DUSTY00  & 2000 & 3.5 & 0.51 &   1.35e-3   &1.85e-5 & 1.0 & 0.006 & [*]\\
  LOri168 & NextGen  & 2900 & 4.0 & 0.45 &   3.99e-4   &7.44e-6 &  & & [2]\\
  LOri169 & NextGen  & 3100 & 4.5 & 0.45 &   2.21e-4   &6.63e-6 &  & & [3]\\
  LOri170 & NextGen  & 3500 & 3.5 & 0.45 &   1.99e-4   &8.21e-6 &  & & [2]\\
\end{longtable}
}


\end{document}